\documentclass[twocolumn,trackchanges]{aastex63}

\usepackage[whole]{bxcjkjatype} 

\newcommand{\kms}{km\,s$^{-1}$}
\newcommand{\ergs}{erg\,s$^{-1}$}

\received{June 22, 2020}
\accepted{December 19, 2020}
\submitjournal{ApJ}

\shorttitle{Sz 102 [\ion{Ne}{3}] Jet from HST/STIS}
\shortauthors{Liu et al.}

\begin{document}

\title{Revealing Ionization Conditions of Sz 102 with Spatially Resolved [\ion{Ne}{3}] Microjets}

\email{cfliu@asiaa.sinica.edu.tw, shang@asiaa.sinica.edu.tw}

\author[0000-0002-1624-6545]{Chun-Fan Liu（劉君帆）}
\affiliation{Institute of Astronomy and Astrophysics, Academia Sinica, Taipei 10617, Taiwan}

\author[0000-0001-8385-9838]{Hsien Shang（尚賢）}
\affiliation{Institute of Astronomy and Astrophysics, Academia Sinica, Taipei 10617, Taiwan}

\author[0000-0002-7154-6065]{Gregory J. Herczeg（沈雷歌）}
\affiliation{The Kavli Institute for Astronomy and Astrophysics, Peking University, Beijing 100871, China}

\author[0000-0001-7796-1756]{Frederick M. Walter}
\affiliation{Department of Physics and Astronomy, Stony Brook University, Stony Brook, NY 11794-3800, USA}

\date{\today}

\begin{abstract}

Forbidden neon emission lines from small-scale microjets can probe high-energy processes in low-mass young stellar systems. We obtained spatially resolved [\ion{Ne}{3}] spectra of the microjets from the classical T Tauri Star Sz 102 using the {\it Hubble Space Telescope Imaging Spectrograph} ({\it HST}/STIS) at a spatial resolution of $\sim0\farcs1$. The blueshifted and redshifted [\ion{Ne}{3}] emission both peak in intensity within $\sim0\farcs1$ of the star and gradually decay along the flow outward to $\sim 0\farcs24$. The spatial distribution and extent of the [\ion{Ne}{3}] microjet is consistent with a jet that is ionized close to the base and subsequently recombines on a longer timescale than the flow time. \ion{Ca}{2} H and K lines are also detected from the redshifted microjet with a line full-width at half-maximum of $\sim170$ \kms, consistent with those of other forbidden emission lines, atop a 300-\kms\ wide stellar component.
The launching radius of the Sz 102 jet, inferred from the observed line centroids and the range of inclination angles and stellar masses from the literature, is on the order of $\sim0.03$ au.
The possible proximity of the launching region to the star allows immediate ionization without distance dilution from the circumstellar ionization sources, most likely keV X-ray flares generated by magnetic reconnection events in the star--disk system, to sustain the observed [\ion{Ne}{3}] flux.

\end{abstract}

\keywords{ISM: individual objects (Sz 102) --- ISM: jets and outflows --- ISM: kinematics and dynamics --- stars:
mass-loss --- stars: pre-main sequence --- X-rays: stars}

\section{Introduction} \label{sec:intro}

In the current star-formation paradigm, the protostar is actively accreting gas from the magnetic field lines that truncate the disk inner edge and channel material from the disk onto the star \citep{XWind_IV,Muzerolle2001,AG12}.
The angular momentum is carried away by jets and outflows launched from the inner sub-au scales of the disk, in the form of magnetocentrifugal mechanism as in \citet{BP82}, although the launch physics and the range of positions for the winds vary among models \citep{Shu2000_PPIV,FC13}. Understanding how jets and winds are launched from the inner disk is an important key missing piece of physics in elucidating how gas accretes through the disk and onto the star.

Studying the jet-launching region requires sub-au resolution observations. Emission lines in the jet help to trace the physical conditions in the launch region \citep[e.g.,][]{SGSL,Panoglou2012}
by spatially resolving the structure of the arcsecond-scale ``microjets''
close to the star \citep[e.g.,][]{Bacciotti2000,Woitas2002,Coffey2008,LS12}
and by line ratio diagnostics using species with various critical densities
\citep{BE99,HM07}.
The observations of optical jets have demonstrated that those jets are often partially ionized, but the origin of this ionization is not well established. Understanding the origin of ionization and excitation in the jet leads to important insights into the physical environment in the wind-launching regions.

The physical properties of neon can help constrain the main ionization source of protostellar jets.
The major channels of neon ionization are through ejection of the outer (L) shell and inner (K) shell electrons. Photon energies in the extreme ultraviolet (EUV) are required to overcome the first two L-shell ionization thresholds of 21.6 eV and 41.0 eV \citep{HG09}. 
KeV X-rays can exceed the $0.87$ and $0.88$ keV K-shell thresholds of \ion{Ne}{1} and \ion{Ne}{2} to produce neon of higher ionization states through the Auger process \citep{GNI07,Muller2017}. 
Although EUV photons are capable of producing ionization up to \ion{Ne}{5}, large cross sections of charge transfer with hydrogen rapidly converts back to its singly and doubly ionized states \citep{GNI07}. Sustained ionization to the higher ionization states is therefore required in the physical processes of generating emission.

Several fine-structure and forbidden lines of [\ion{Ne}{2}] (Ne$^+$) and [\ion{Ne}{3}] (Ne$^{2+}$) are seen toward low-mass young stellar objects (YSOs). 
[\ion{Ne}{2}]~12.81 \micron\ has been detected in more than 50 low-mass YSOs in the {\it Spitzer}/IRS survey, suggesting the ubiquitous existence of warm ($\sim 500$ K) circumstellar gas irradiated by X-ray and EUV photons \citep{Pascucci2007,Lahuis2007,Guedel2010,Baldovin-Saavedra2011,Espaillat2013}.
High-dispersion mid-infrared spectroscopy of [\ion{Ne}{2}] was used to decipher possible mechanisms from the line profiles, including disk atmosphere, photoevaporative winds, and jets and outflows \citep[e.g.,][]{Alexander2008,Herczeg2007,Pascucci2009,Najita2009,Baldovin-Saavedra2012,Sacco2012}.
\citet{SGLL} predicted that neon emission lines would arise from the innermost region of the jets where ample high energy photons can penetrate, and suggested strong correlations can exist between the neon line luminosities and products of the total X-ray luminosity and mass-loss rate ($L_{\rm X}\dot{M}_{\rm w}$). \citet{Guedel2010} found that the [\ion{Ne}{2}] luminosity correlates with $L_{\rm X}\dot{M}_{\rm w}$, and that the
[\ion{Ne}{2}] luminosity is 1--2 orders of magnitude larger in jet-driving YSOs than in other YSOs, where it correlates with the [\ion{O}{1}] luminosity, a mass-loss indicator.

Detections of [\ion{Ne}{3}] emission from low-mass YSOs are much less common. Only 5 low-mass YSOs have been reported to emit the fine-structure [\ion{Ne}{3}] 15.55\micron\ line \citep{Pascucci2007,Lahuis2007,Espaillat2013}.
The forbidden [\ion{Ne}{3}]~$\lambda\lambda3869, 3967$ lines are specifically sensitive to partially ionized gas up to $\sim 10^4$ K at critical densities of $\sim10^7$ cm$^{-3}$ and can serve as tracers for jets surrounding the high-energy environment \citep{SGLL}. 
To date [\ion{Ne}{3}]~$\lambda\lambda3869, 3967$ has been seen only in the microjets of Sz~102 \citep{Liu14}, DG~Tau \citep{Liu16}, and ESO-H$\alpha$~574 \citep{Whelan2014}.

Various scenarios of neon ionization are proposed to account for the detections of forbidden neon emission lines. In accreting star--disk systems, EUV and soft (sub-keV) X-ray photons can be produced in the accretion shocks on the stellar surface with a shock velocity of several hundred km\,s$^{-1}$ \citep{Kastner2002,Schneider2018}. 
Harder keV X-rays can be produced in magnetic reconnection events arising in star--disk systems \citep{SSGL}.
For T Tauri stars with mass accretion rates larger than $10^{-8}$ $M_\odot\,{\rm yr}^{-1}$, the circumstellar material may have absorbed the majority of the EUV and soft X-ray photons nonetheless, leaving mainly the hard X-rays to penetrate through wind to reach disk atmosphere \citep{GNI04,HG09,Pascucci2014}. Likewise, the hard X-rays dominate jet ionization from within. 
Magnetic flares can heat up the plasma up to $T_{\rm X} \sim 100$ MK and emit hard X-rays \citep{PF02}, with luminosities $\gtrsim 10^{31}$ erg\,s$^{-1}$ \citep[see, e.g.,][]{Favata2005,Wolk2005}.
Alternatively, collisional ionization can produce [\ion{Ne}{3}] when shock speeds are faster than $\sim 100$ km\,s$^{-1}$ \citep{HG09}, which is found near the terminal bow shocks where the outflows strongly interact with the ambient material \citep[e.g., HH34;][]{Morse1993}, and potentially in very bright knots.
The DG Tau jet may be an example of strong jet shocks \citep[e.g.,][]{SS08,Schneider2013,Guedel2008,Guedel2011}, although the formation of a $> 100$ km\,s$^{-1}$ internal jet shock is not fully understood \citep[see also][]{Guenther2009,Guenther2014}.

In this work, we further examine the physical conditions in the innermost regions of a YSO using spatially resolved {\it Hubble Space Telescope Imaging Spectrograph} ({\it HST}/STIS) spectra of the Sz 102 microjet on a scale of $\sim0\farcs1$ ($\sim16$ au), taking advantage of its edge-on geometry (the target, Sz 102, is briefly described in Sec.\ \ref{subsec:sz102_neon}). With a spatially resolved variation in [\ion{Ne}{3}] and other emission lines, a distinction between different major ionization scenarios could be identified.

We briefly summarize the properties of previous neon line detections and the latest {\it HST}/STIS spectra of Sz 102 microjet in Sections \ref{sec:sz102_HST}. We present the results of the spatially resolved {\it HST}/STIS spectra in \ref{sec:results}, including forbidden [\ion{Ne}{3}], [\ion{O}{2}], and [\ion{S}{2}] and permitted \ion{Ca}{2} emission lines. Based on the spatial distribution of [\ion{Ne}{3}]$\lambda3869$ emission, we discuss the possibility that Sz 102 is predominantly ionized at the base of the jet in Section \ref{sec:discussion}, and summarize this work in Section \ref{sec:summary}.

\section{The HST/STIS Spectra of Sz 102} \label{sec:sz102_HST}

\subsection{Previous Neon Detections of Sz 102} \label{subsec:sz102_neon}

Sz 102, a highly veiled K-type star in Lupus III, is one of the few young stars with both [\ion{Ne}{2}]~12.81 \micron\ and [\ion{Ne}{3}]~15.55 \micron\ detections \citep{Lahuis2007,Espaillat2013}. 
Its [\ion{Ne}{2}] flux is among the highest in the $\sim50$ {\it Spitzer}/IRS stellar samples \citep{Guedel2010}, although the emission was not captured in a high-dispersion spectroscopic observation \citep{Pascucci2009}. 
Sz 102 drives bipolar jets that are viewed close to the plane of sky and extend up to $5\arcsec$ in both optical and near-infrared emission lines \citep{BE99,Coffey2010}. \citet{Liu14} discovered [\ion{Ne}{3}]~$\lambda3869$ within $\sim 200$ au ($1\farcs3$ at 160~pc; see Sec. \ref{subsec:sz102_properties}) of star in high-dispersion ($R\approx33,000$) VLT/{\sc Uves} spectra.
The spatially unresolved (seeing $\sim1\farcs5$) emission was spectroastrometrically decomposed into redshifted and blueshifted components separated by $\sim 0\farcs3$. Each side of the microjets has a line width larger than $\sim140$ \kms\ (FWHM), which suggests origination in a wide-angle wind.

\subsection{Observation and Data Reduction of the HST/STIS Spectra} \label{subsec:obs}

Sz 102 and its bipolar microjets were observed with {\it HST}/STIS on 2016 May 6, under the General Observing (GO) Program 14177 (P.I.: C.-F. Liu).
The plate scale of the STIS CCD is $0\farcs05$/pixel, resulting in an effective spatial resolution of 2 pixels, or $\sim 0\farcs1$.
The {\tt G430M} grating and the {\tt 52X0.2} slit were used throughout the observations. The slit was placed along the jet axis, at a position angle of $98^\circ$ \citep{WH09}. 
This setup yields a dispersion of 0.276 \AA/pixel and a spectral resolution of $R\sim7,000$, or equivalently, a velocity sampling of $\sim21$ km\,s$^{-1}$ and a velocity resolution of $\sim45$ km\,s$^{-1}$ at $\sim 4000$\AA.
Two settings of central wavelengths ({\tt CENWAVE}), 3843 and 3936 \AA, were chosen to cover the 3700 to 3986 \AA\ and 3793 to 4079 \AA\ regions, respectively. Forbidden lines such as [\ion{O}{2}] $\lambda\lambda3726+2729$, [\ion{Ne}{3}] $\lambda3869$, and [\ion{S}{2}] $\lambda4068$ ([\ion{S}{2}] $\lambda4076$, located at the edge of the detector, is unmeasurable) and permitted lines including \ion{Ca}{2} H + K, and Balmer series up to H$\epsilon$ were covered. For each central wavelength setting, two dither positions separated by $0\farcs525$ ($\sim 10$ pixels) along the slit axis were used for further cleansing of bad pixels and cosmic ray hits. The exposure times for {\tt CENWAVE=3843} and {\tt CENWAVE=3936} are 1462 and 1044 seconds, respectively. In all, four exposures were obtained over two orbits in one visit.

We downloaded the pipeline-calibrated STIS {\tt X2D} (two-dimensional spectral imaging) files from Mikulski
Archive for Space Telescopes (MAST). 
These data have been wavelength-rectified and flux-calibrated using CALSTIS v.3.4. Removal of hot/bad pixels and cosmic-ray hits was done in two steps using {\tt CCDProc} \citep{ccdproc_v1.2.0_2016}. 
First the cosmic rays were rejected using a median filter across each of the four images with a $3\sigma$ clipping and a moving box of 7 pixels. After shifting and rectifying the images, we combined the images using a median filter with iterative clipping at $2\sigma$.
The reduced two-dimensional spectral image was sliced into subimages around emission line regions of interest. The subimages were further decomposed in velocity by two-Gaussian fitting along each spatial pixel row with the continuum subtracted using pySpecKit \citep{pyspeckit} as in \citet{Liu14}. 
The resulting spectral subimages were smoothed by convolving with a two-dimensional Gaussian kernel of $\sigma_G = 0.8$ pixels.

\subsection{Adopted Distance and Systemic Velocity of Sz 102} \label{subsec:sz102_properties}

Sz 102 is a member of the Lupus III cloud \citep{Comeron2008_Lupus}.
We adopt a distance of 160 pc to Sz 102, consistent with the mean distance to the Lupus III cloud \citep[e.g.][]{Long2017,Alcala2019} and that adopted by \citet{Fang2018}.
Due to obscuration by its highly inclined disk, the astrometry, proper motion, and distance to the optical counterpart of Sz 102 are highly uncertain. The Gaia DR2 parallax measurement for Sz 102 has a large fractional error and excess astrometric noise and is therefore unreliable.

The velocity of the Lupus III cloud relative to the local standard of rest (LSR) was measured to be $+4$ \kms\ \citep{Tachihara2001,CF10} or $+5$ \kms\ \citep{Coffey2008}. Recent studies of the CO disk of Sz 102 suggest a value of $+3.0\pm0.1$ \kms\ \citep{Louvet2016}, consistent with previous measurements. 
In light of moderate velocity resolution of the STIS data and for compatibility with our previous work \citep{Liu14}, we adopt a systemic velocity of $+4$ \kms; all velocities discussed below are relative to this systemic velocity.

\section{Results} \label{sec:results}

\subsection{One-Dimensional Spectrum} \label{subsec:1dspec}

\begin{figure*}
\plotone{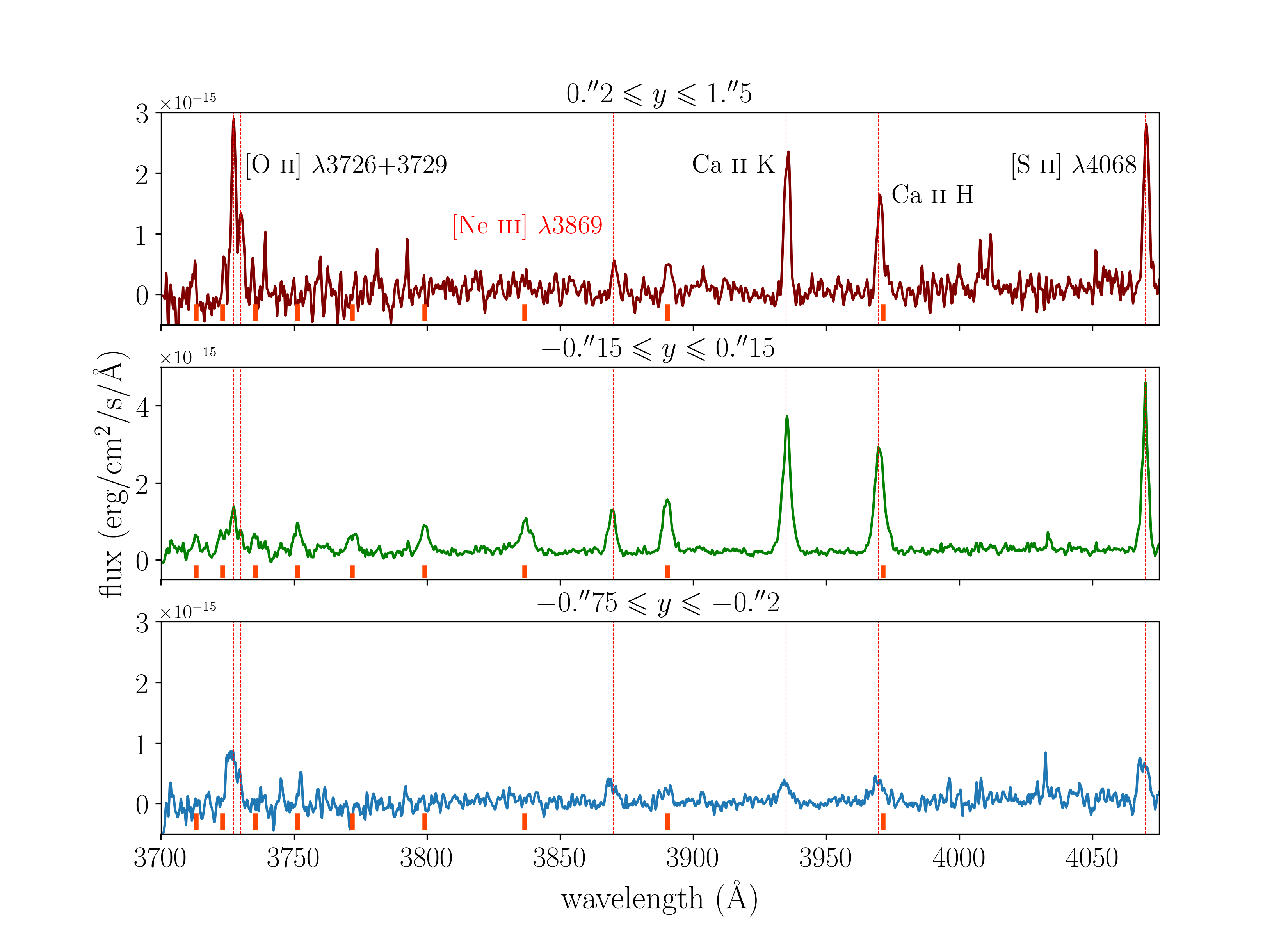}
\caption{Combined one-dimensional spectrum of Sz 102 obtained with two central wavelength settings of the {\it HST}/STIS {\tt G430M} grating. Bright emission lines are labeled with species and wavelengths; the positions of Balmer series from H$\epsilon$ down to H15 are indicated by short bars. \label{fig:STIS_1d_spec}}
\end{figure*}

Utilizing the high spatial resolution of {\it HST}/STIS, we examine the properties of the emission lines at various spatial positions.
Figure \ref{fig:STIS_1d_spec} shows the reduced and combined one-dimension spectra of Sz 102 microjets, before continuum subtraction, integrated over various spatial intervals. The one-dimensional spectra are integrated over $\pm0\farcs15$ from the star (center panel) along with extended emission covering between $-0\farcs75$ and $-0\farcs2$ (the lower panel) and between $0\farcs2$ and $1\farcs5$ (the upper panel). Table \ref{tab:1dspec} shows the line properties at the three aforementioned positions, obtained by Gaussian fitting to the available line centroids.

The line properties change with position on the jet axis. Emission from $y > 0\farcs2$ exhibits velocity centroids ranging between $+30$ to $+40$ \kms, tracing the redshifted microjet. Emission from $y < -0\farcs2$ corresponds to the blueshifted microjet, with velocity centroids around $-40$ to $-50$ \kms. The redshifted microjet shows an overall line width of $160$ to $180$ \kms, and the blueshifted microjet shows a line width of $\sim 350$ \kms. Line properties from close to the star ($|y|< 0\farcs2$) appear to differ among various lines. For the forbidden emission lines, the low-density [\ion{O}{2}] lines have line centroids somewhat redshifted with line widths of $\sim 150$ \kms, whereas the high-density [\ion{Ne}{3}] and [\ion{S}{2}] lines are blueshifted with line widths of $\sim 200$ \kms. On the other hand, the permitted \ion{Ca}{2} H + K doublet lines are slightly redshifted with large line widths of $\sim 300$ \kms.

The relative line strengths between emission lines vary along the slit. Close to the star, the \ion{Ca}{2} H + K doublet lines are the strongest permitted lines; Balmer lines from H$\epsilon$ (blended with \ion{Ca}{2} H) down to H15 are detected toward the star. [\ion{S}{2}] $\lambda4068$ is the strongest among all emission lines and both [\ion{O}{2}] $\lambda\lambda3726+3729$ and [\ion{Ne}{3}] $\lambda3869$ are detected at comparable line fluxes. At distances further away from the star, the relative strengths of the lines change. In the region spatially corresponding to the redshifted microjet, the [\ion{O}{2}] lines become much stronger and the combined flux exceeds those of [\ion{Ne}{3}] $\lambda3869$ and [\ion{S}{2}] $\lambda4068$. The \ion{Ca}{2} H and K lines remain bright in the redshifted microjet, whereas the Balmer lines become undetected except for possible detections of H$\epsilon$ and H$\eta$. All the forbidden emission lines are detected toward the blueshifted microjet, whereas \ion{Ca}{2} H and K lines are the only detectable permitted lines. The spatial dependence of the emission line intensities will be further presented and discussed in the next subsection (Sec. \ref{subsec:2dspec}).

\begin{deluxetable*}{ccccccccccccc}
\tablecaption{Properties of Emission Lines at Various Positions \label{tab:1dspec}}
\tabletypesize{\footnotesize}
\tablewidth{\textwidth}
\tablehead{$y_{\rm range}$ (\arcsec) & $\lambda_{\rm cent}$ (\AA) & $v_{\rm cent}$ (km\,s$^{-1}$) & $v_{\rm width}$ (km\,s$^{-1}$) & flux (erg\,s$^{-1}$\,cm$^{-2}$)}
\startdata
 \multicolumn{5}{c}{[\ion{O}{2}] $\lambda3726$} \\
 $0\farcs2<y<1\farcs5$ & $3727.3\pm0.2$ & $11.2\pm16.5$ & $155\pm32$ & $6.04\times10^{-15}$ \\
 $-0\farcs15<y<0\farcs15$ & $3727.2\pm0.3$ & $6.64\pm23.1$ & $174\pm64$ & $2.48\times10^{-15}$ \\
 $-0\farcs75<y<-0\farcs2$ & $3726.6\pm0.3$ & $-42.1\pm25.2$ & $385\pm83$ & $4.39\times10^{-15}$ \\
\hline \hline
 \multicolumn{5}{c}{[\ion{O}{2}] $\lambda3729$} \\
 $0\farcs2<y<1\farcs5$ & $3730.1\pm0.2$ & $11.2\pm16.4$ & $181\pm93$ & $3.61\times10^{-15}$ \\
 $-0\farcs15<y<0\farcs15$ & $3730.0\pm0.3$ & $6.63\pm23.1$ & $118\pm77$ & $8.35\times10^{-16}$ \\
 $-0\farcs75<y<-0\farcs2$ & $3729.4\pm0.3$ & $-42.1\pm25.1$ & $43\pm62$ & $2.52\times10^{-16}$ \\
\hline \hline
 \multicolumn{5}{c}{[\ion{Ne}{3}] $\lambda3869$} \\
 $0\farcs2<y<1\farcs5$ & $3870.4\pm0.7$ & $40.0\pm57.8$ & $160\pm131$ & $1.08\times10^{-15}$ \\
 $-0\farcs15<y<0\farcs15$ & $3869.5\pm0.3$ & $-32.3\pm19.8$ & $236\pm45$ & $3.45\times10^{-15}$ \\
 $-0\farcs75<y<-0\farcs2$ & $3869.2\pm0.9$ & $-56.1\pm71.9$ & $266\pm158$ & $1.18\times10^{-15}$ \\
\hline \hline
 \multicolumn{5}{c}{\ion{Ca}{2} K} \\
 $0\farcs2<y<1\farcs5$ & $3935.2\pm0.2$ & $30.1\pm12.8$ & $185\pm28$ & $6.04\times10^{-15}$ \\
 $-0\farcs15<y<0\farcs15$ & $3935.1\pm0.1$ & $18.6\pm10.0$ & $305\pm21$ & $1.21\times10^{-14}$ \\
 $-0\farcs75<y<-0\farcs2$ & $3934.3\pm0.9$ & $-40.6\pm65.3$ & $343\pm153$ & $1.57\times10^{-15}$ \\
\hline \hline
 \multicolumn{5}{c}{\ion{Ca}{2} H} \\
 $0\farcs2<y<1\farcs5$ & $3970.2\pm0.2$ & $41.2\pm14.9$ & $197\pm37$ & $4.27\times10^{-15}$ \\
 $-0\farcs15<y<0\farcs15$ & $3969.8\pm2.4$ & $8.25\pm178$ & $308\pm116$ & $1.08\times10^{-14}$ \\
 $-0\farcs75<y<-0\farcs2$ & $3969.3\pm0.9$ & $-24.2\pm66.2$ & $377\pm161$ & $1.69\times10^{-15}$ \\
\hline \hline
 \multicolumn{5}{c}{[\ion{S}{2}] $\lambda4068$} \\
 $0\farcs2<y<1\farcs5$ & $4070.2\pm0.1$ & $26.4\pm9.78$ & $171\pm18$ & $6.50\times10^{-15}$ \\
 $-0\farcs15<y<0\farcs15$ & $4069.7\pm0.1$ & $-5.6\pm3.6$ & $202\pm6$ & $1.06\times10^{-14}$ \\
 $-0\farcs75<y<-0\farcs2$ & $4069.2\pm0.3$ & $-42.4\pm23.0$ & $339\pm60$ & $3.97\times10^{-15}$ \\
\enddata
\end{deluxetable*}

\subsection{Spatially-Resolved Two-Dimensional Spectral Decomposition} \label{subsec:2dspec}

\begin{figure*}
\gridline{\fig{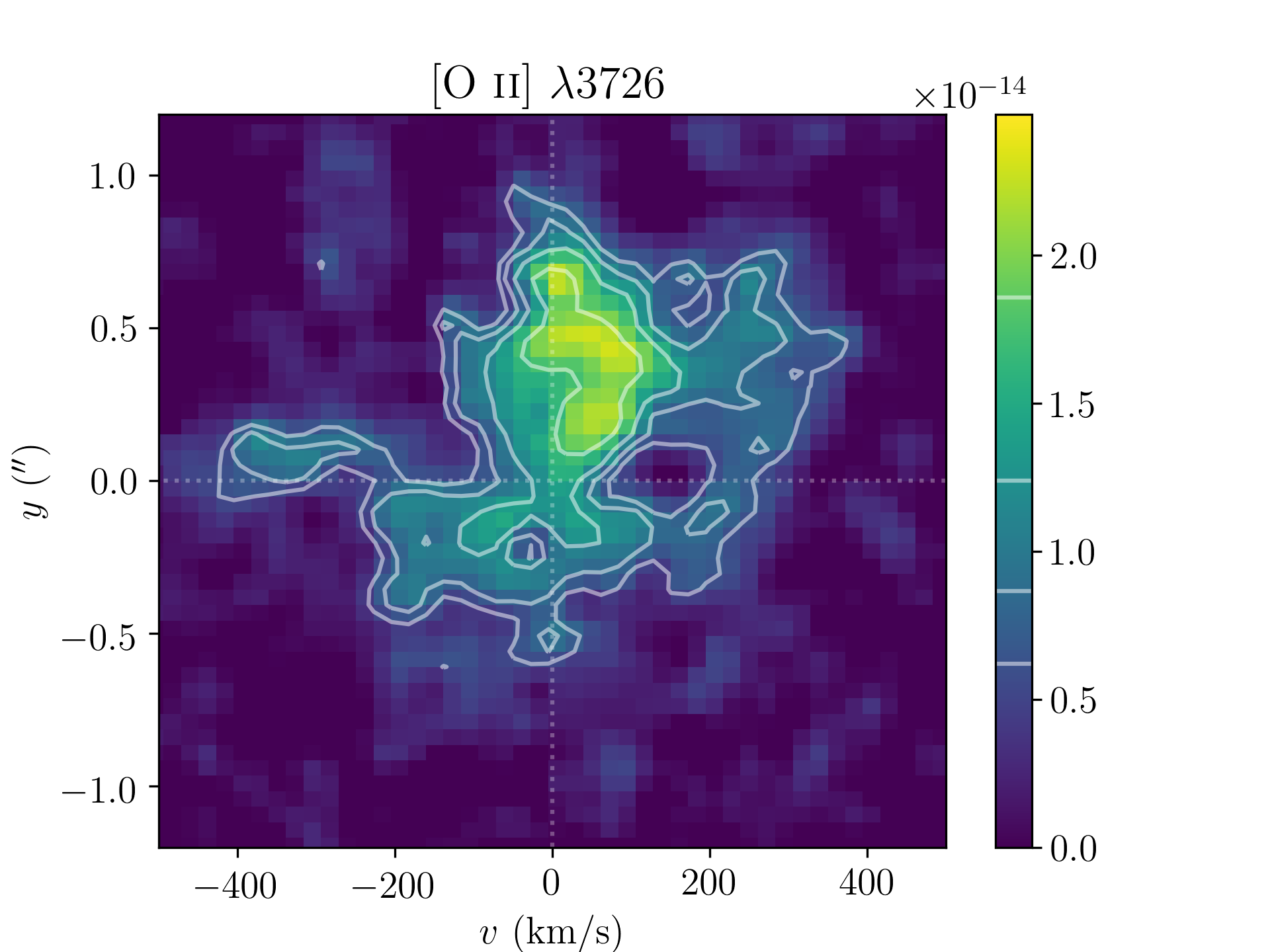}{0.5\textwidth}{(a)}
          \fig{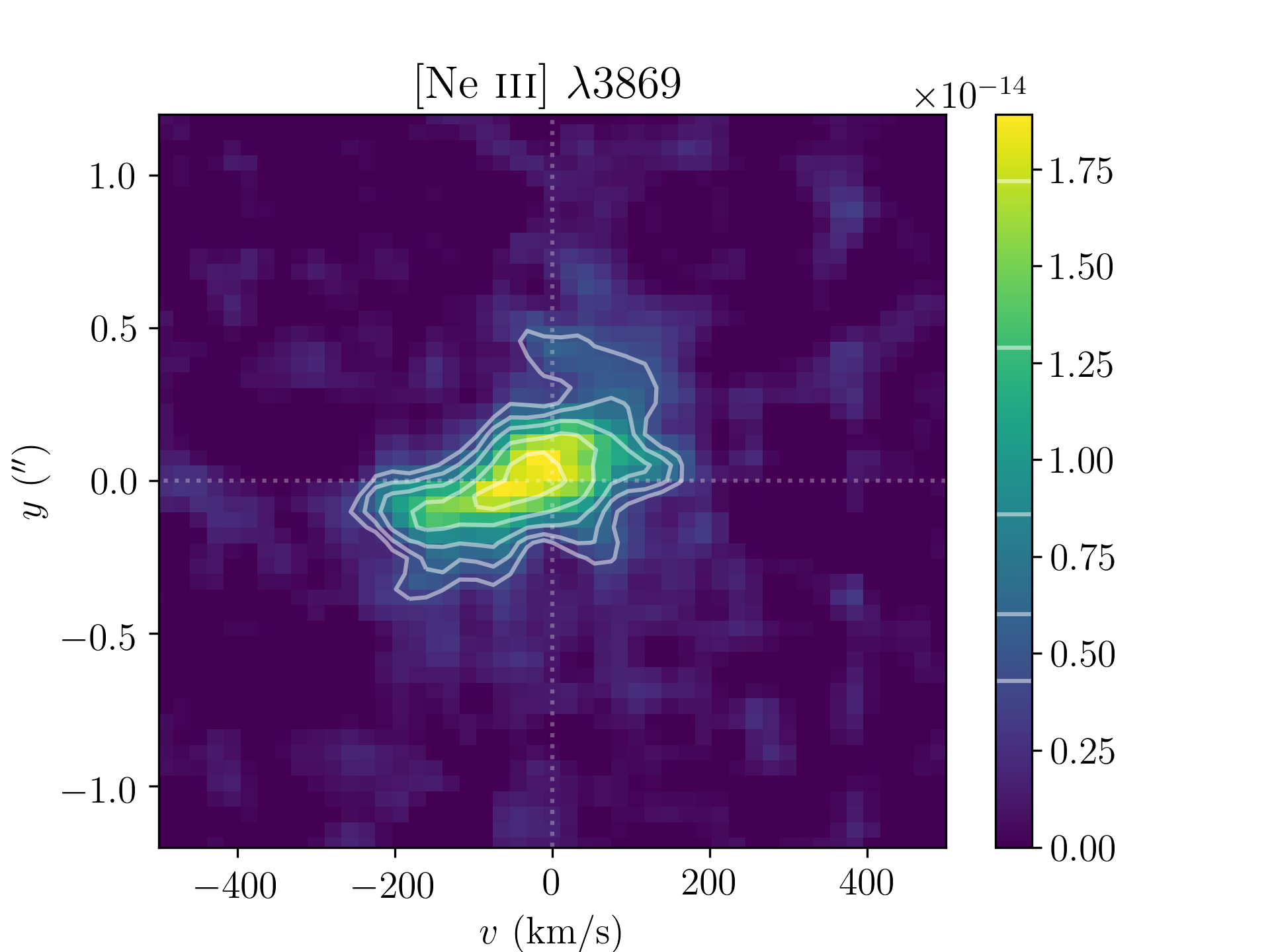}{0.5\textwidth}{(b)} 
         }
\gridline{\fig{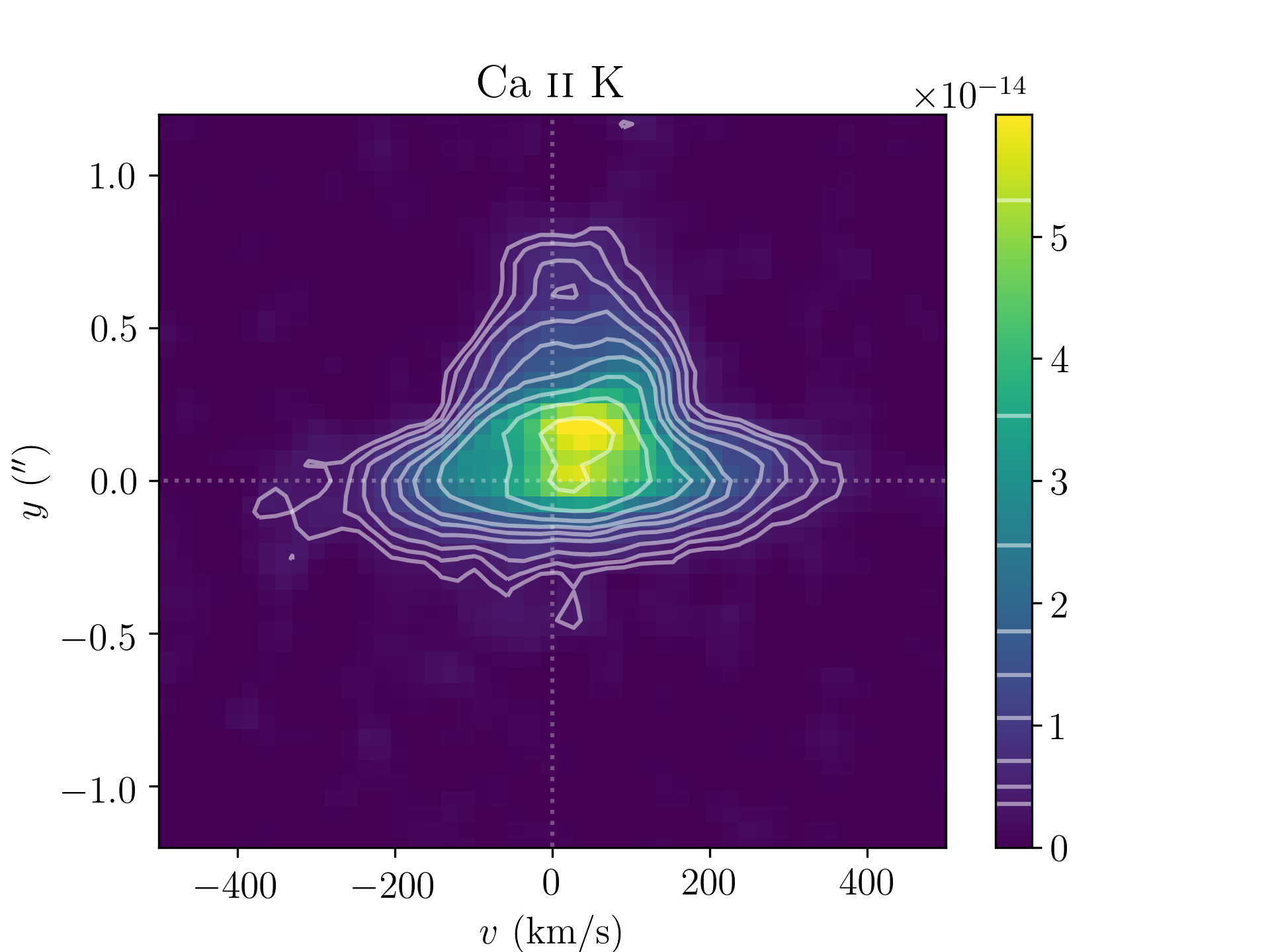}{0.5\textwidth}{(c)}
          \fig{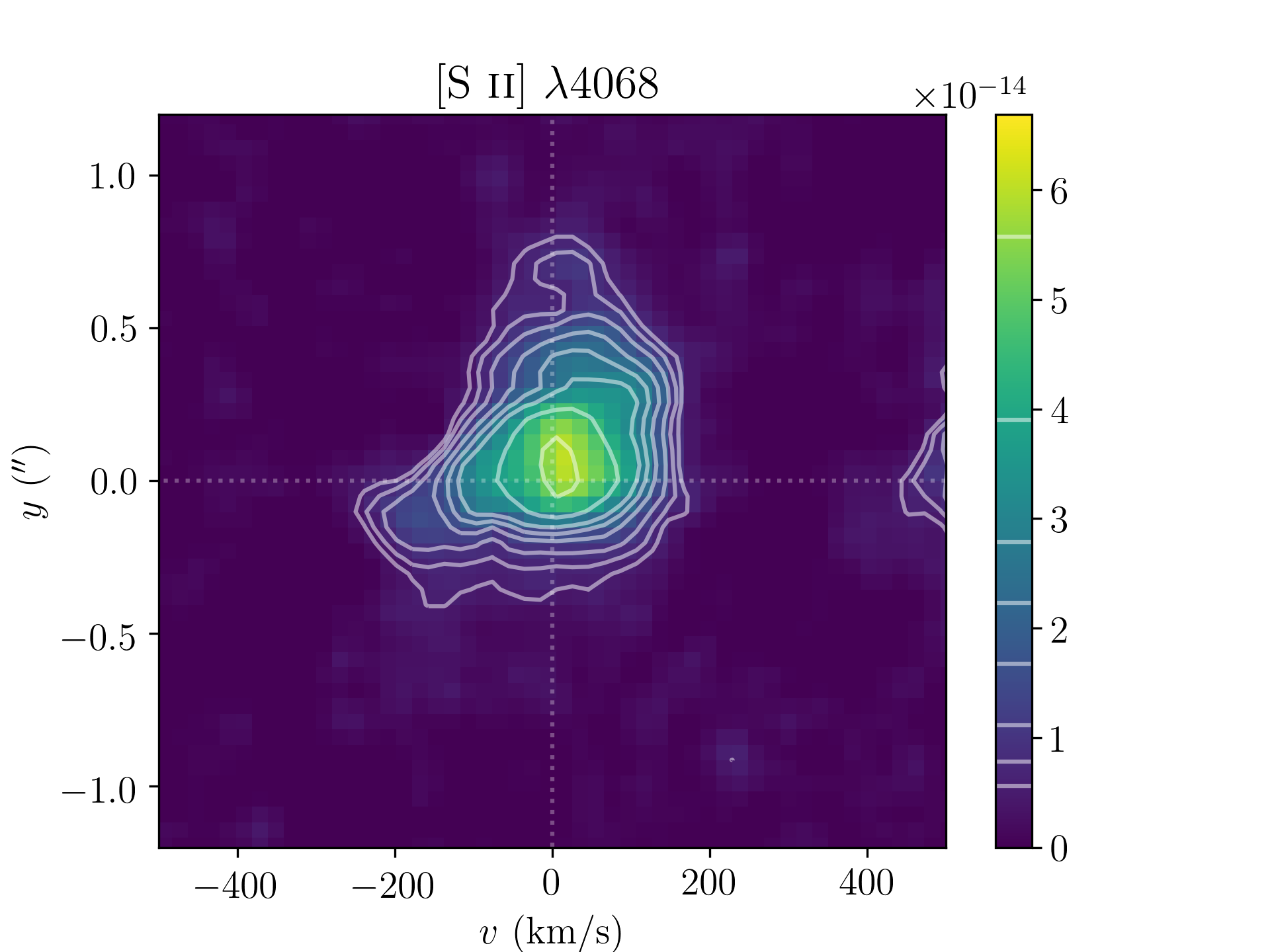}{0.5\textwidth}{(d)}
         }
\caption{Two-dimensional spectral images in the form of position--velocity diagrams at the wavelengths of (a) [\ion{Ne}{3}] $\lambda3869$, (b) [\ion{O}{2}] $\lambda3726+3729$ (the blob at $\sim-350$ \kms\ is H14), (c) \ion{Ca}{2} K, and (d) [\ion{S}{2}] $\lambda4068$. The white contours show intensity in $5$, $7$, $10$, $15$, $20$, $25$, $35$, $50$$\sigma$, where $\sigma$ is the rms values of each line: $8.60\times10^{-16}$ for [\ion{Ne}{3}], $1.24\times10^{-15}$ for [\ion{O}{2}], $7.06\times10^{-16}$ for \ion{Ca}{2} K, and $1.11\times10^{-15}$ for [\ion{S}{2}], in units of erg\,cm$^{-2}$\,s$^{-1}$\,\AA$^{-1}$\,arcsec$^{-2}$. \label{fig:STIS_2d_spec}}
\end{figure*}

Figure \ref{fig:STIS_2d_spec} shows the position--velocity (pv) diagrams of the bright emission lines obtained from the calibrated two-dimensional spectral images. From the pv diagrams, the first noticeable feature is the spatial extension of the forbidden lines along the jet axis. The line emission extends from close to the star down to $\sim0\farcs5$ in the blueshifted jet and to $\sim1\farcs0$ in the redshifted jet. For the [\ion{Ne}{3}] and [\ion{S}{2}] lines, with a high critical densities ($n_{\rm cr} \approx 10^6$ cm$^{-3}$), the intensity peaks within $\sim0\farcs1$ and fades as the jet propagates. For [\ion{O}{2}] emission, with its critical density $\sim 3$ orders of magnitude lower, the peak position is somewhat further down the jet, at $\sim0\farcs4$ in the redshifted jet.
A remarkable finding from the pv diagrams is that the \ion{Ca}{2} H and K lines are also traced down the redshifted microjet. The pv diagram of \ion{Ca}{2} shows that the emission at the star has a large line width of $\sim 300$ \kms, whereas the extended redshifted microjet has a line width of $\sim 180$ \kms. There might be blueshifted \ion{Ca}{2} emission as suggested by the spectrum $\sim0\farcs5$ offset from the star (Figure \ref{fig:STIS_1d_spec}) and the marginal extension in the pv diagram but its properties are difficult to extract because it is much fainter than the broad stellar component.

\begin{figure*}
\gridline{\fig{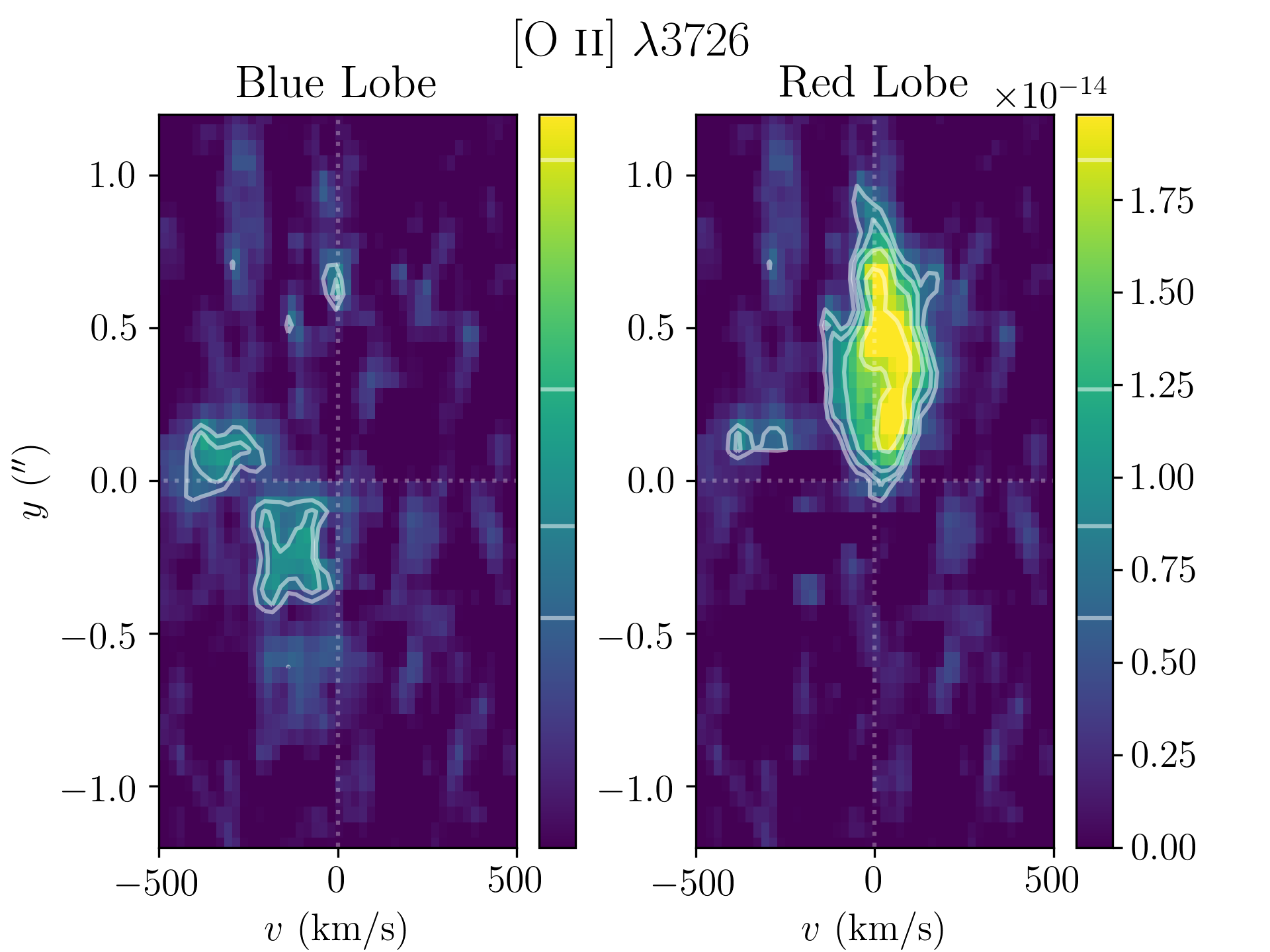}{0.5\textwidth}{(a)}
          \fig{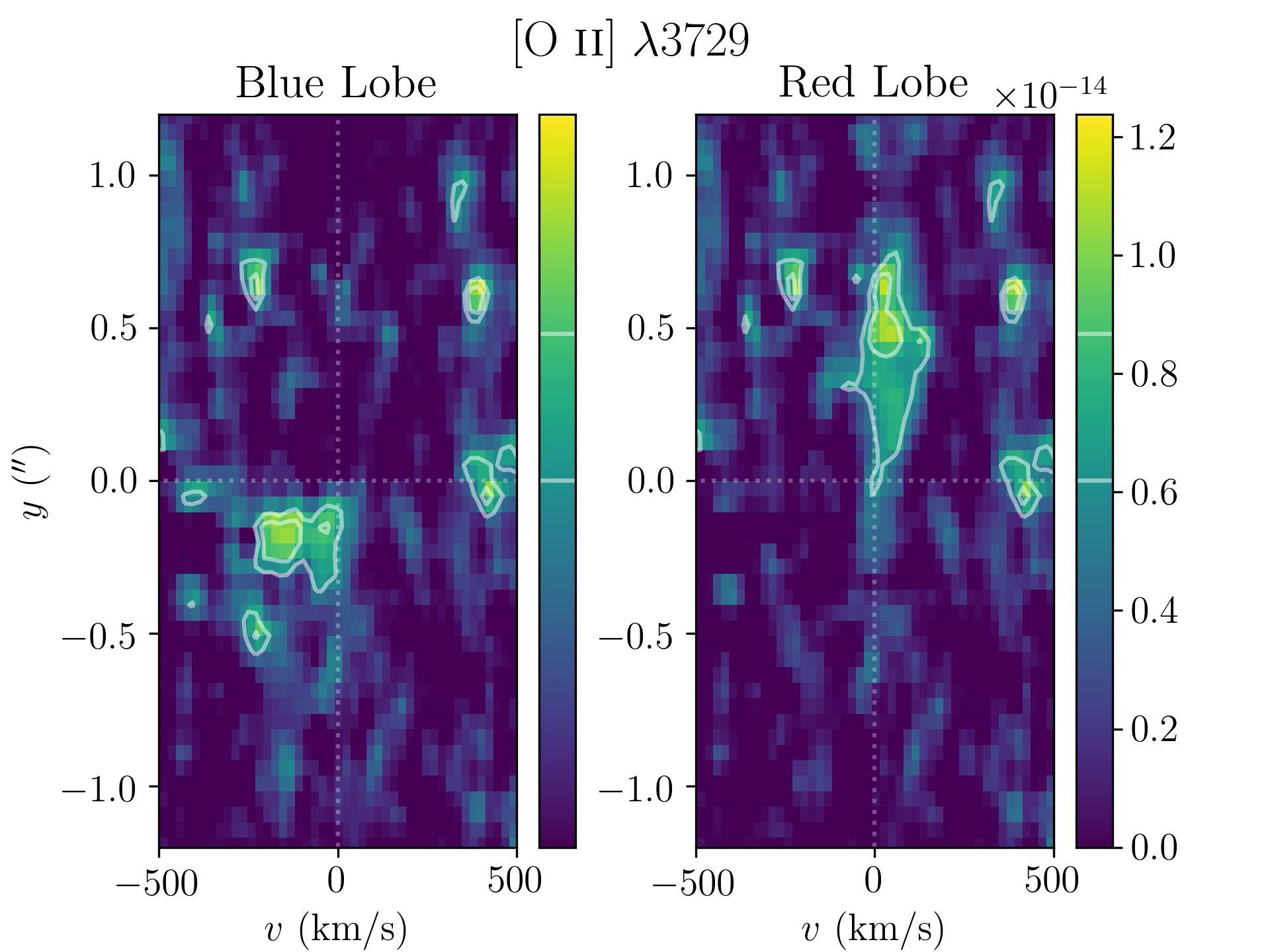}{0.5\textwidth}{(b)}
         }
\gridline{\fig{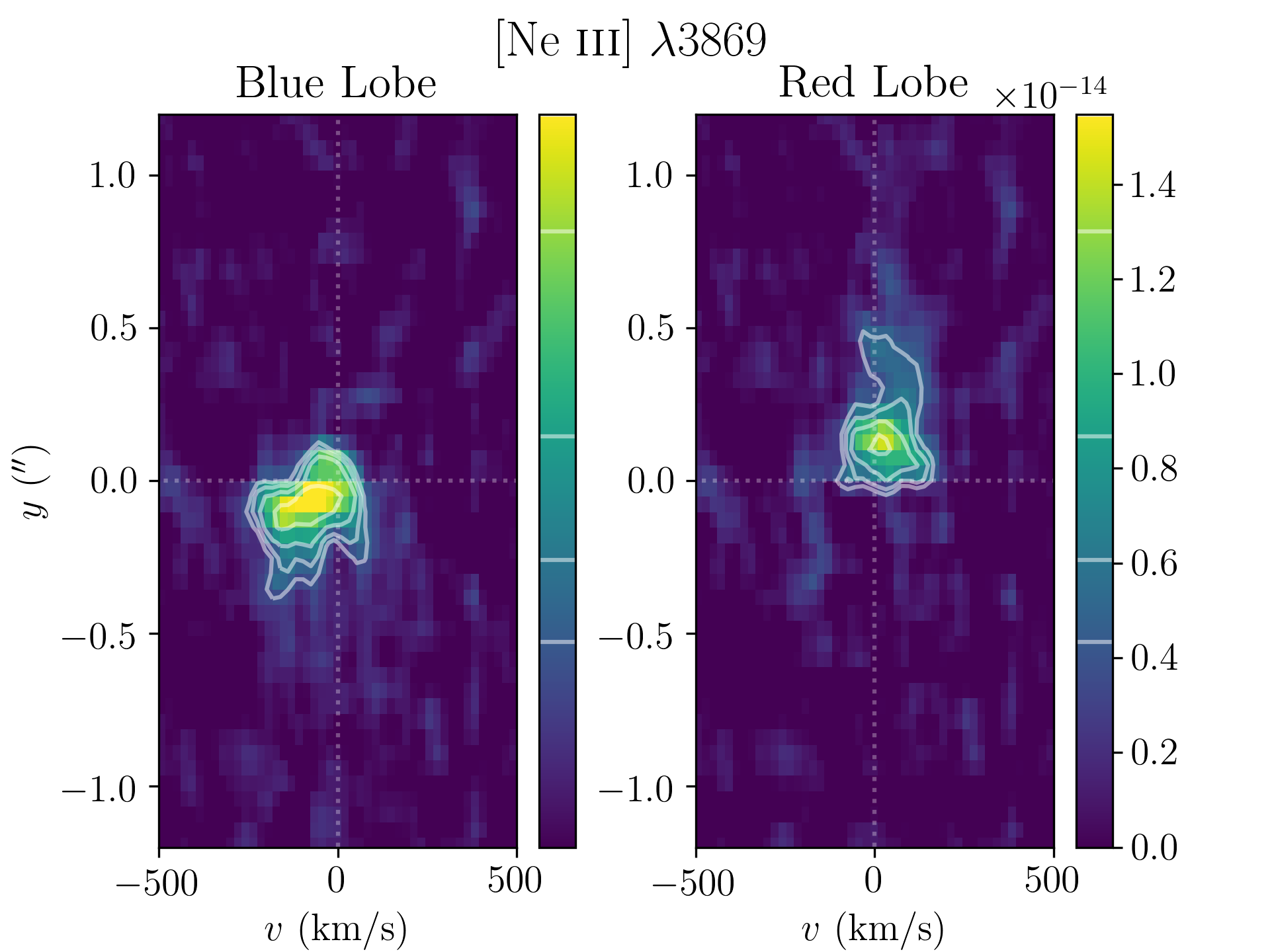}{0.5\textwidth}{(c)}
          \fig{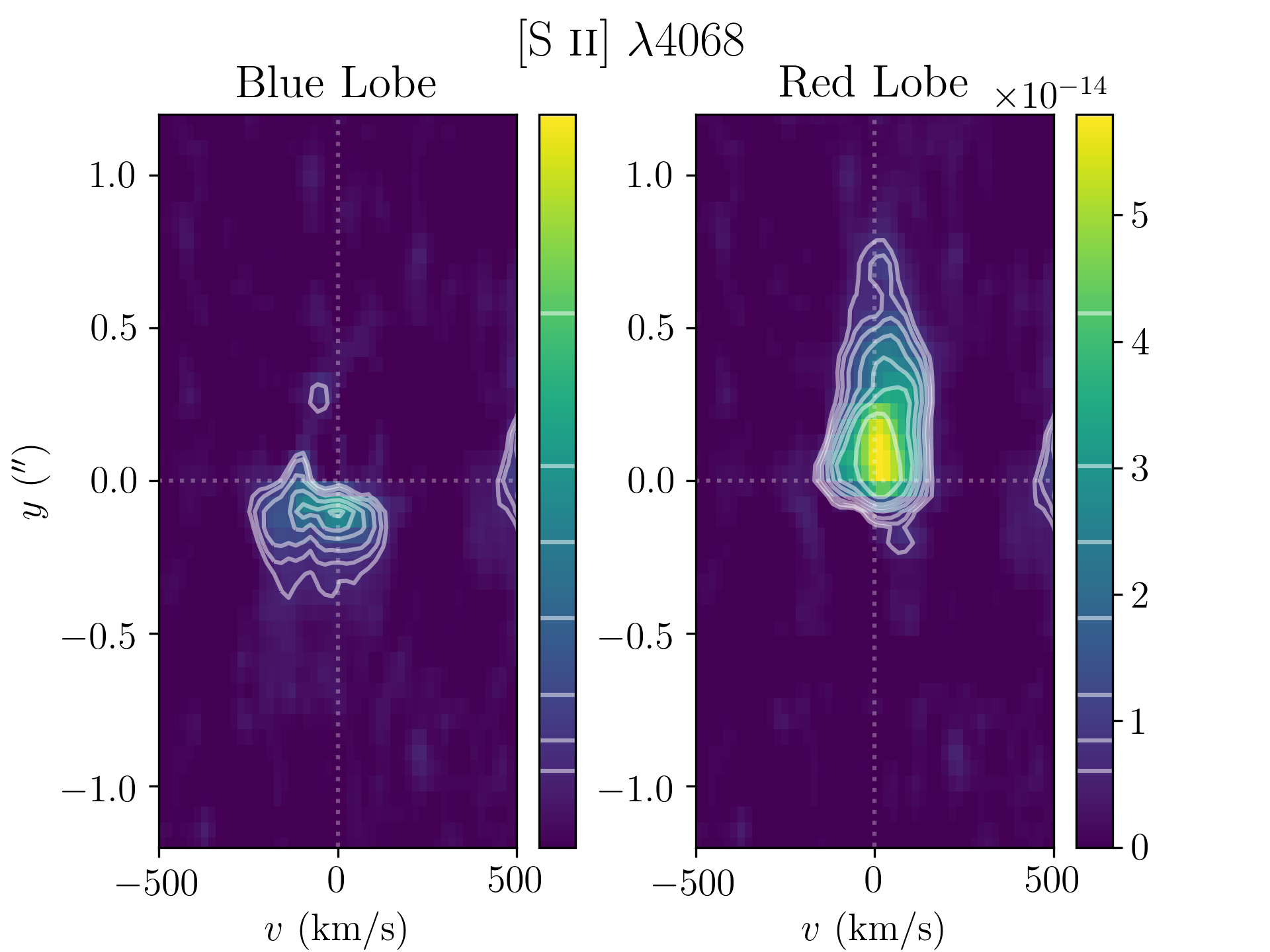}{0.5\textwidth}{(d)}
         }
\caption{Two-dimensional spectral images in the form of position--velocity diagrams at the wavelengths of (a) [\ion{O}{2}] $\lambda3726$ (the blob at $\sim-350$ \kms\ is H14), (b) [\ion{O}{2}] $\lambda3729$, (c) [\ion{Ne}{3}] $\lambda3869$, and (d) [\ion{S}{2}] $\lambda4068$. For each of the panels, the left side and the right side shows the blueshifted and redshifted microjet component after velocity decomposition. The white contours show intensity in $5$, $7$, $10$, $15$, $20$, $25$, $35$$\sigma$, where $\sigma$ is the rms values of each line: $8.53\times10^{-16}$ for [\ion{Ne}{3}], $1.25\times10^{-15}$ for [\ion{O}{2}], and $1.12\times10^{-15}$ for [\ion{S}{2}], in units of erg\,cm$^{-2}$\,s$^{-1}$\,\AA$^{-1}$\,arcsec$^{-2}$.} \label{fig:STIS_2d_spec_vdecomp}
\end{figure*}

\begin{figure*}
\gridline{\fig{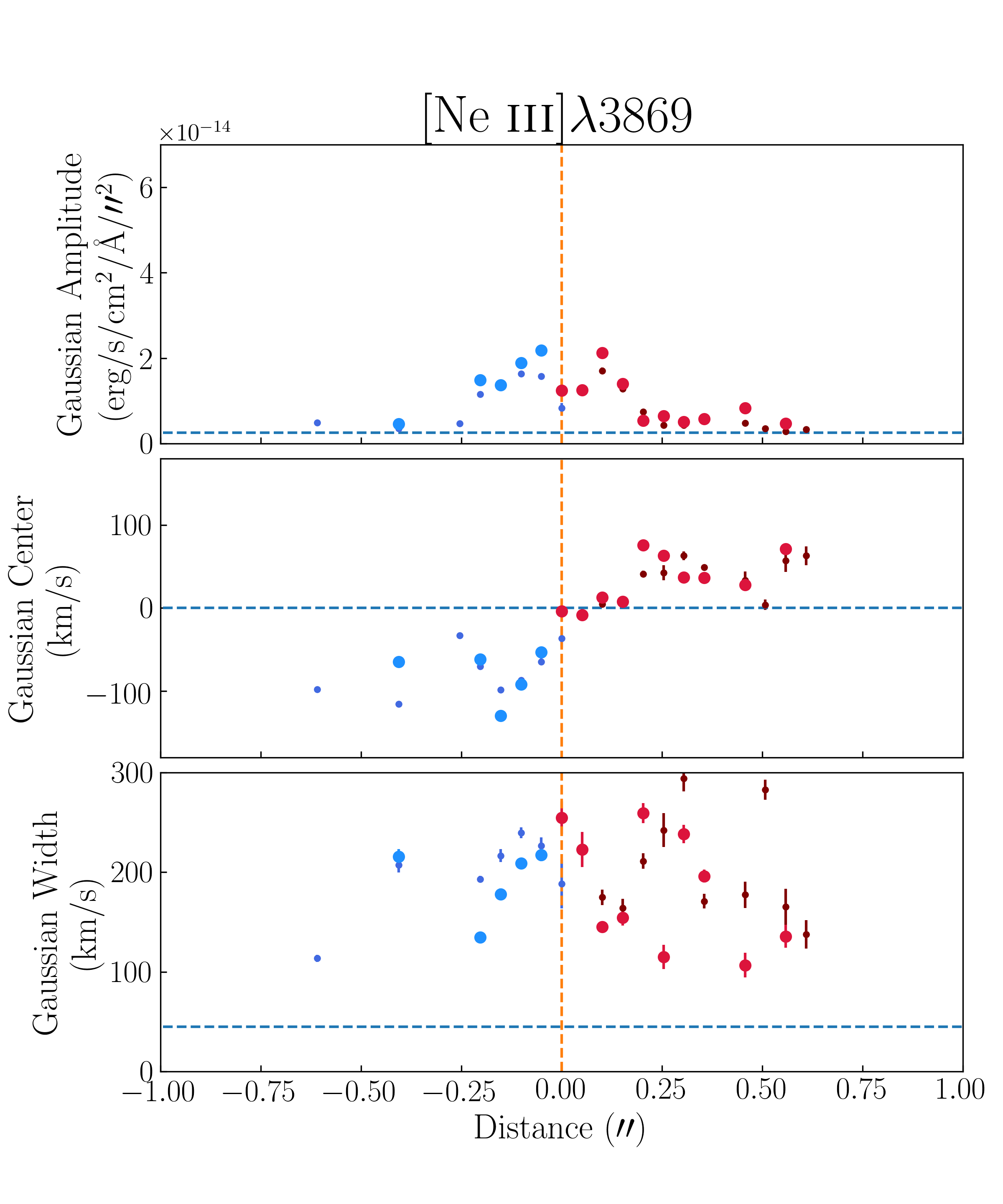}{0.32\textwidth}{(a)}
          \fig{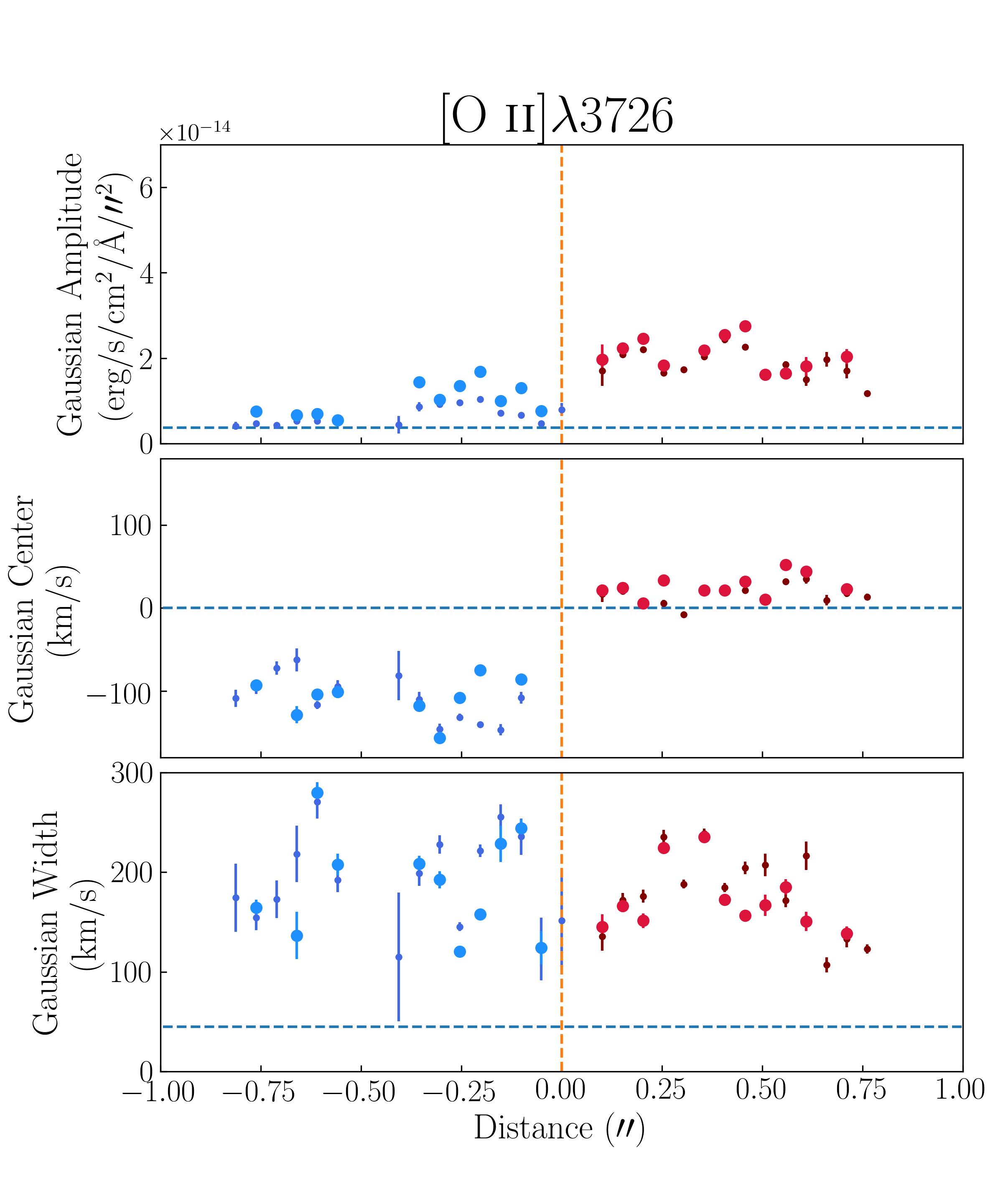}{0.32\textwidth}{(b)}
          \fig{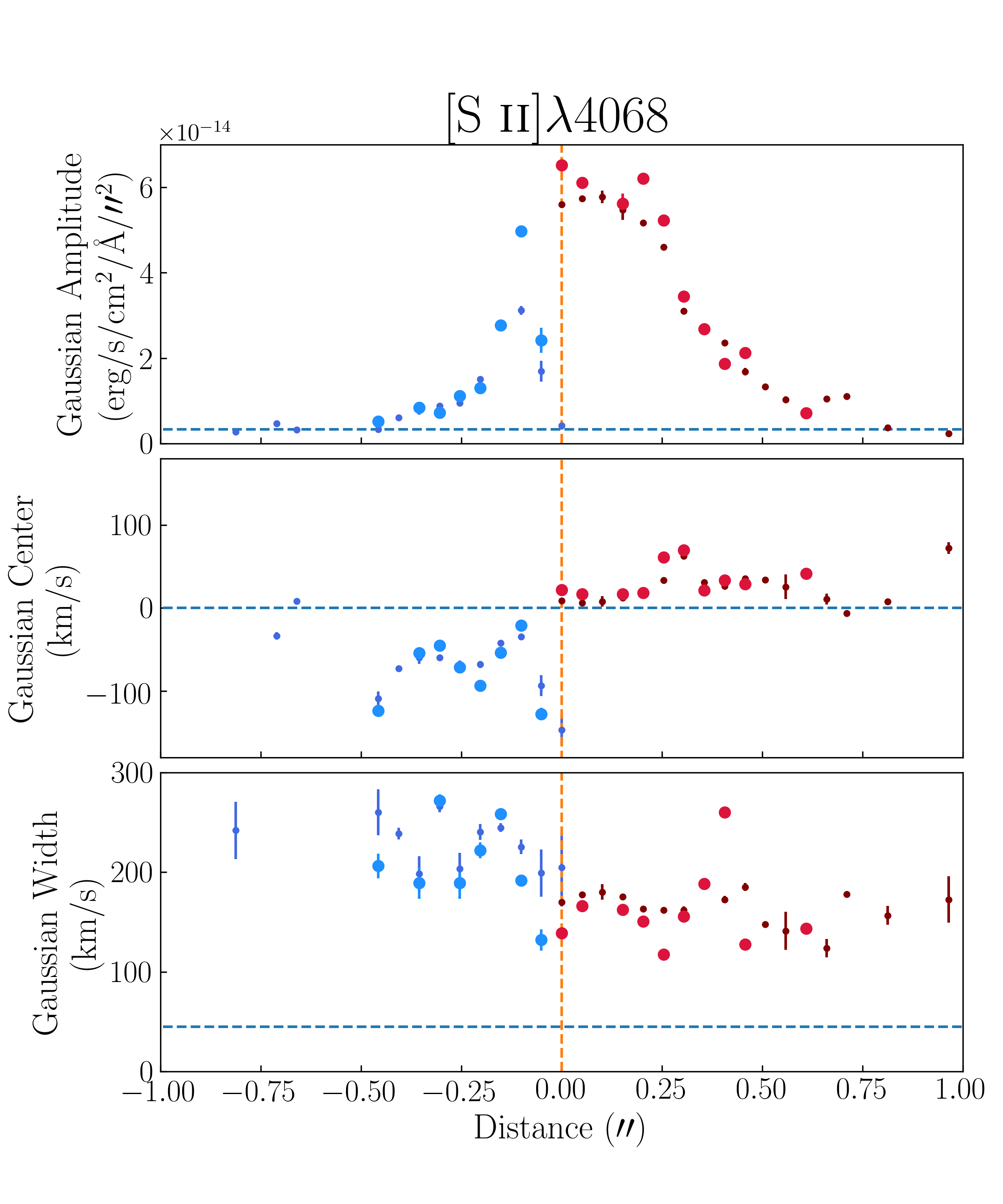}{0.32\textwidth}{(c)}}
\caption{Fitted parameters of blueshifted velocity component (blue symbols) and the redshifted velocity component (red symbols) of three main forbidden emission lines. The smaller dots and bigger filled circles are from reduced data with and without Gaussian smoothing. The horizontal dashed line in the amplitude panel represents $2\sigma$ of the line region. The horizontal dashed line in the line width panel represents the nominal velocity resolution of $\sim45$ \kms.} \label{fig:STIS_velpars_fitting}
\end{figure*}

Properties of the jet emission can be further diagnosed in the velocity-decomposed spectra shown in Figure \ref{fig:STIS_2d_spec_vdecomp}. 
The blueshifted and redshifted microjets show comparable [\ion{Ne}{3}] line fluxes. 
The [\ion{O}{2}] and [\ion{S}{2}] lines show a more dominant redshifted microjet and a low signal-to-noise faint blueshifted microjet. 
Figure \ref{fig:STIS_velpars_fitting} shows the line profile parameters (amplitudes, velocity centroids, and velocity widths) obtained by velocity-decomposition Gaussian fitting along the jet axis. 
The brightness of the redshifted microjet in the higher critical density [\ion{Ne}{3}] and [\ion{S}{2}] lines gradually decreases outwards along the flow.
The intensity peaks emerge at $\sim0\farcs1$ from the center, and fade below $\sim3\sigma$ at $\sim0\farcs25$ for [\ion{Ne}{3}] $\lambda3869$ and at $\sim0\farcs6$ for [\ion{S}{2}] $\lambda4068$. There are no clear signs of detached knots within the innermost $1\farcs0$ of the jets. The [\ion{O}{2}] $\lambda3726$ line does not exhibit an overall intensity peak in the redshifted microjet. 
It appears to extend from $\sim0\farcs1$ up to $\sim0\farcs75$ at similar intensities with minor discontinuities at $\sim0\farcs25$ and $\sim0\farcs5$. All the forbidden emission lines are confined within $\sim0\farcs25$ of the blueshifted microjet. They show a similar trend as the redshifted microjet in that the [\ion{Ne}{3}] and [\ion{S}{2}] lines peak close to the star and fade away along the flow and that the [\ion{O}{2}] lines maintain similar intensity along the flow. The median velocity centroids of the redshifted microjet is $\sim+20$ \kms, and that of the blueshifted microjet is $\sim-70$ \kms.
The trends of line widths along the jets are less conspicuous due to lower signal-to-noise ratios in each spatial position. The redshifted microjet has a median value of $\sim150$ \kms\ whereas the blueshifted microjet is largely scattered around $\sim200$ \kms.

\begin{figure}
\plotone{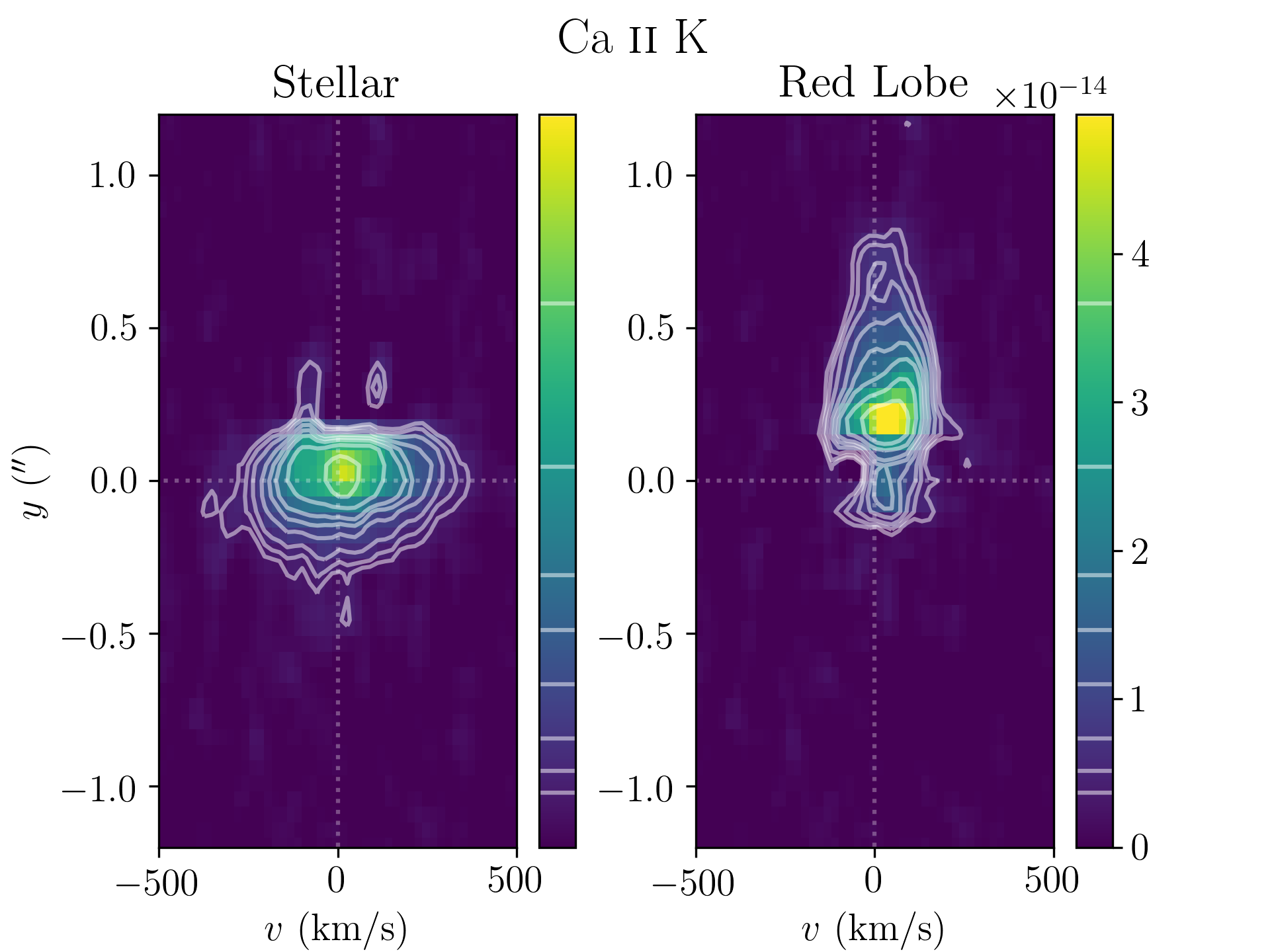}\\
\plotone{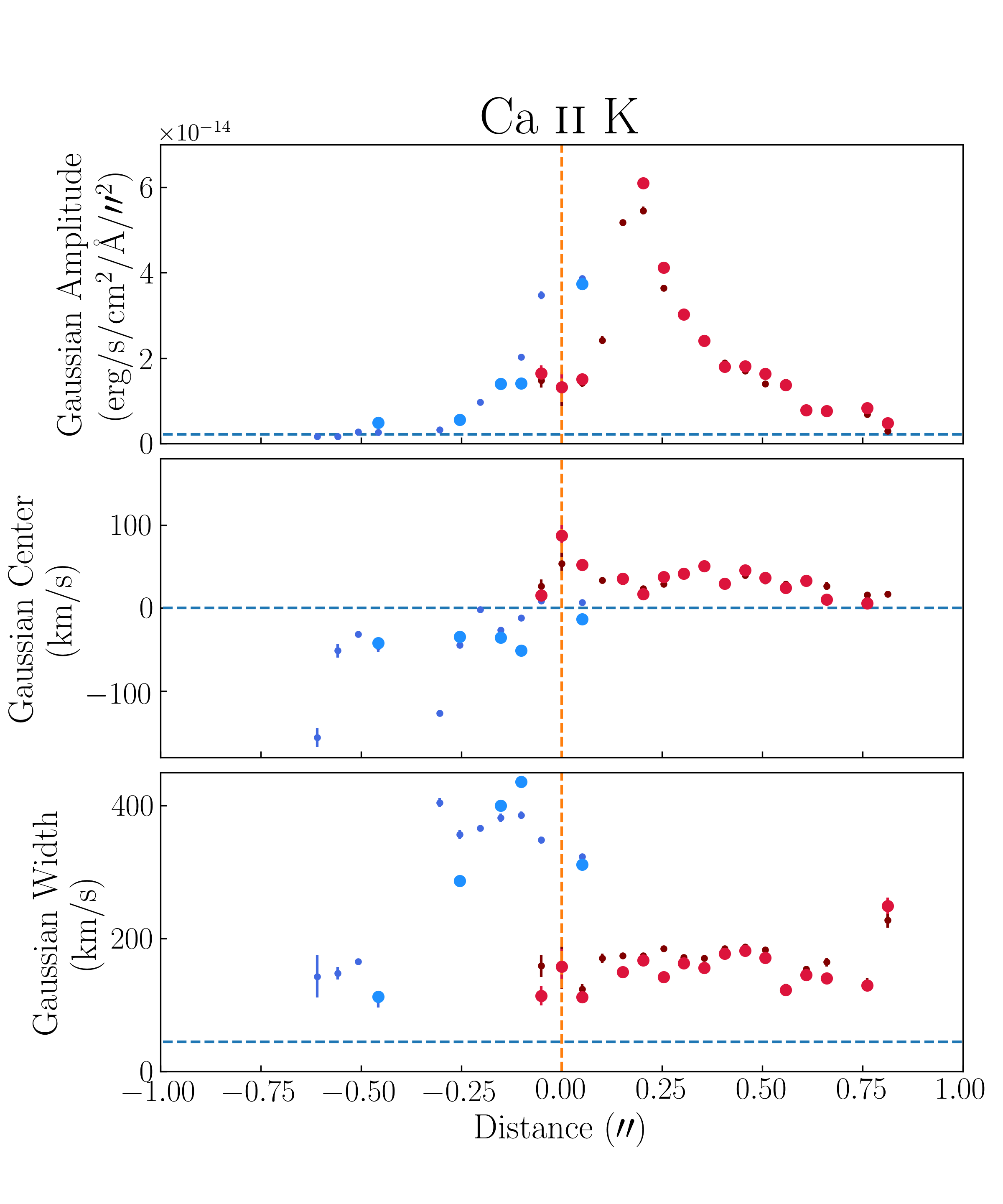}
\caption{Velocity-decomposed position--velocity diagrams (top) and the Gaussian-fitting results (bottom) of \ion{Ca}{2} K emission line, showing the stellar component with a large line width (left) and the redshifted microjet component extended up to $\sim0\farcs5$ (right).} \label{fig:STIS_2d_CaIIK_vdecomp}
\end{figure}

Figure \ref{fig:STIS_2d_CaIIK_vdecomp} shows the velocity-decomposed pv diagram of \ion{Ca}{2} K emission line and the properties of its Gaussian-fitting parameters. The two distinct contributions from the stellar component and redshifted microjet are more evident in the velocity-decomposed pv diagram. The \ion{Ca}{2} K line shows a pattern similar to those of the high-density forbidden lines in the redshifted microjet, with a peak intensity close to the star and fading along the flow. The peak position is somewhat further away from the star, at $\sim0\farcs2$, and the emission extends up to $\sim0\farcs75$. The velocity centroids are $\sim+30$ \kms\ and the line width $\sim150$ \kms, consistent with the properties of the detected forbidden emission lines. The other component, centered at the nominal stellar position and systemic velocity, has a large line width of $\gtrsim350$ \kms. Its large line width and small velocity centroid value differentiates it from the contribution of a microjet. The apparent asymmetric shape in the pv diagram toward the blueshifted emission may be indicative of the existence of a \ion{Ca}{2} blueshifted microjet, but the brighter stellar component hinders one from further investigations into the current dataset.

\begin{figure}
\plotone{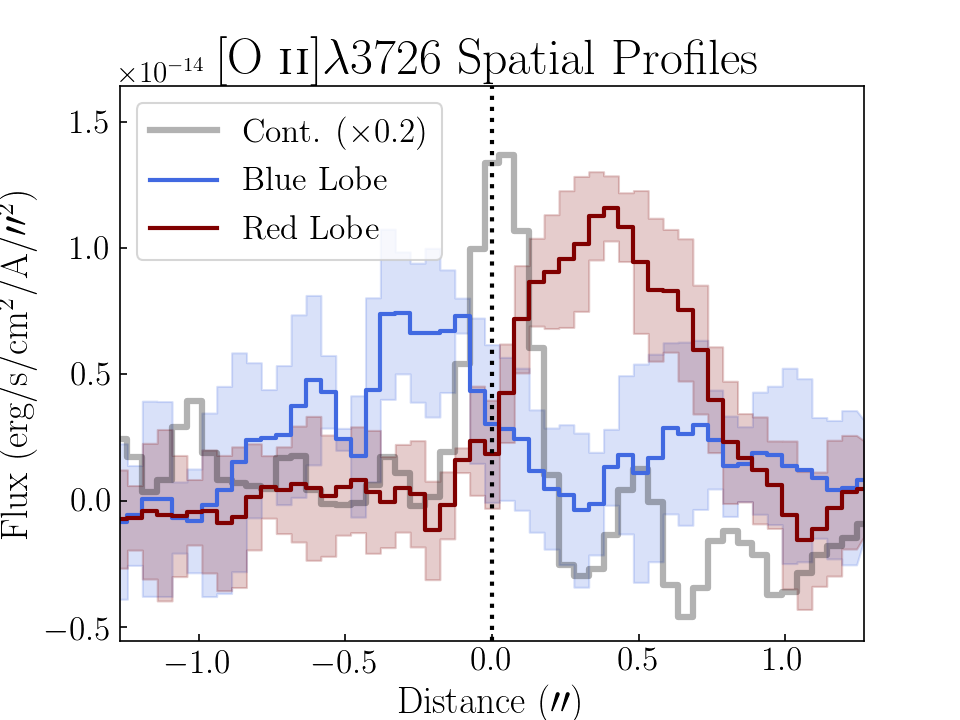} \\
\plotone{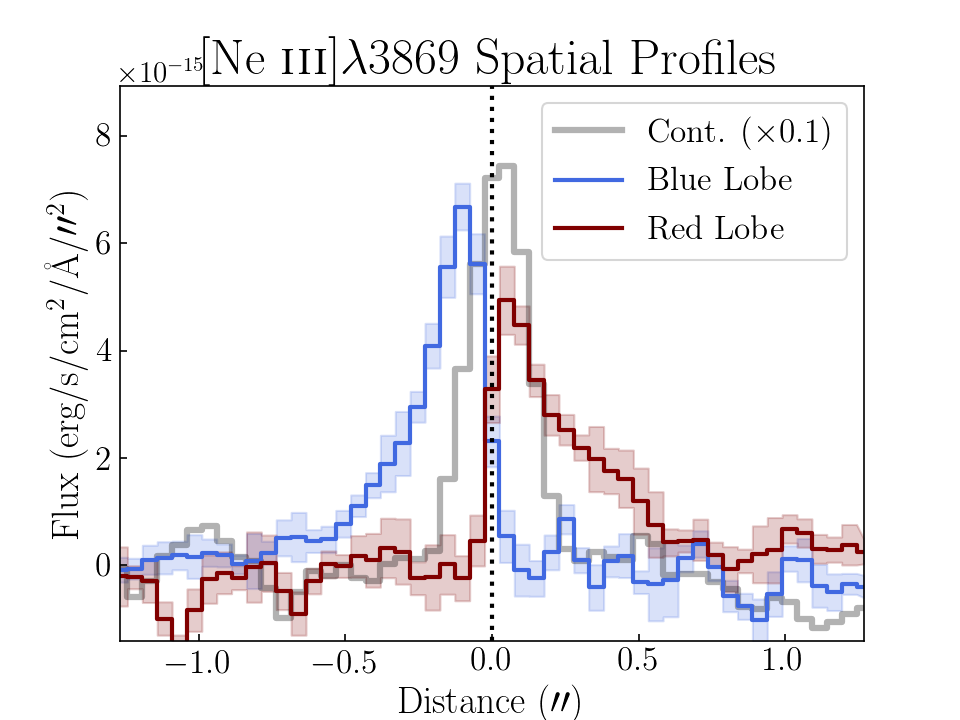} \\
\plotone{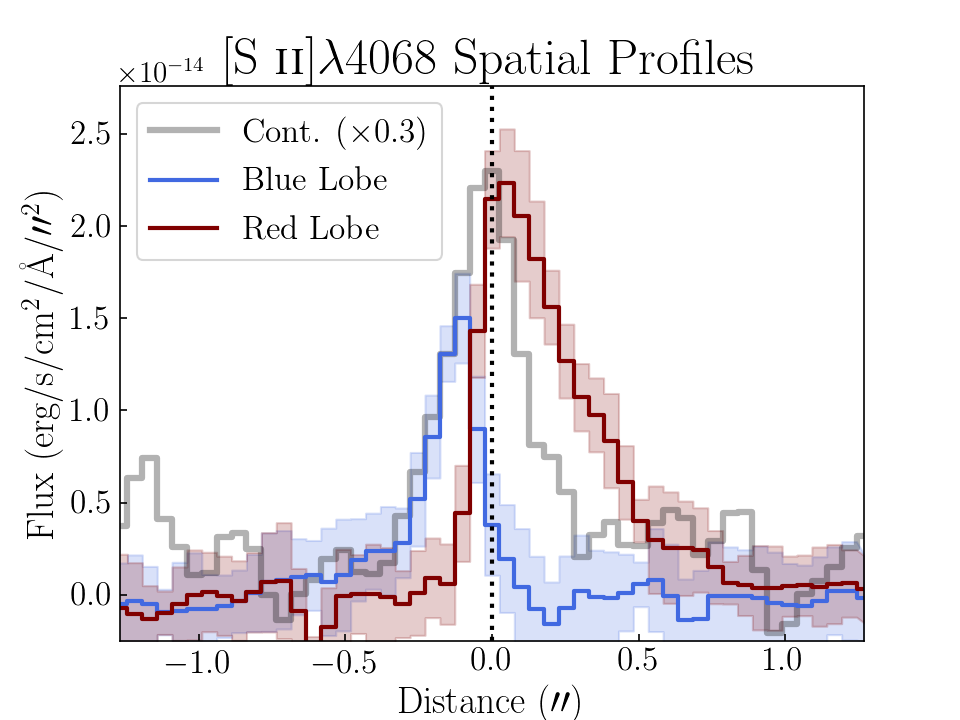}
\caption{Spatial profiles of the blueshifted emission (blue) and those of the redshifted emission (red). The continuum spatial profiles from their adjacent line-free regions are shown in gray, which integrate over [$-900,-600$] \kms\ in [\ion{O}{2}] line, and [$-700,-400$] \kms\ in [\ion{Ne}{3}] and [\ion{S}{2}] lines. The profiles were smoothed with a Gaussian of width $\sigma_G=0.8$ cells for better presentations.} \label{fig:STIS_1d_spatprof}
\end{figure}

\begin{figure}
\plotone{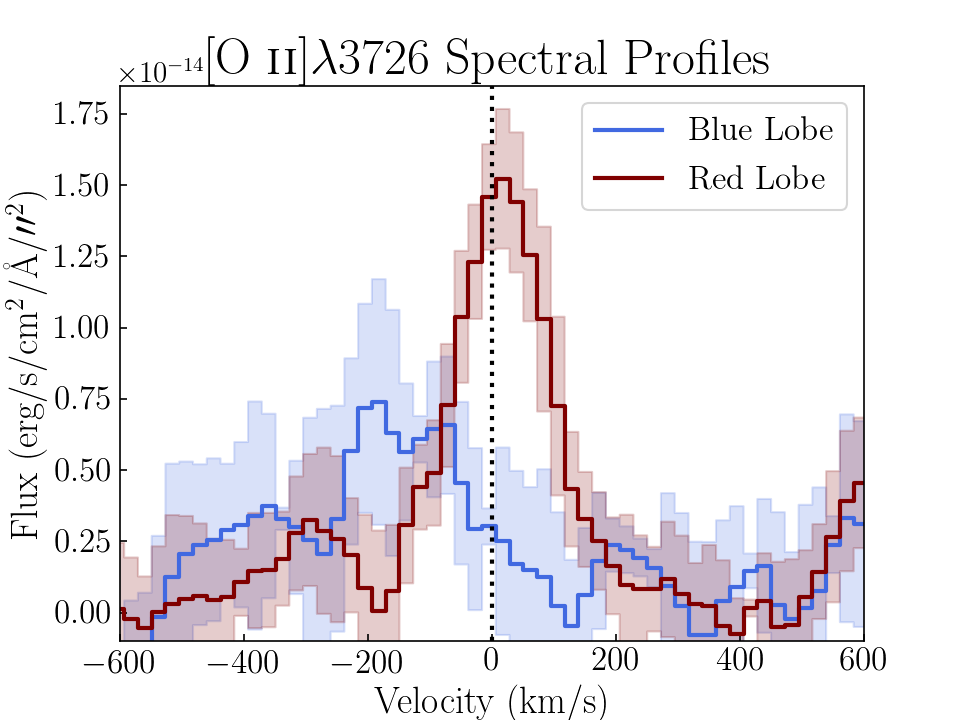} \\
\plotone{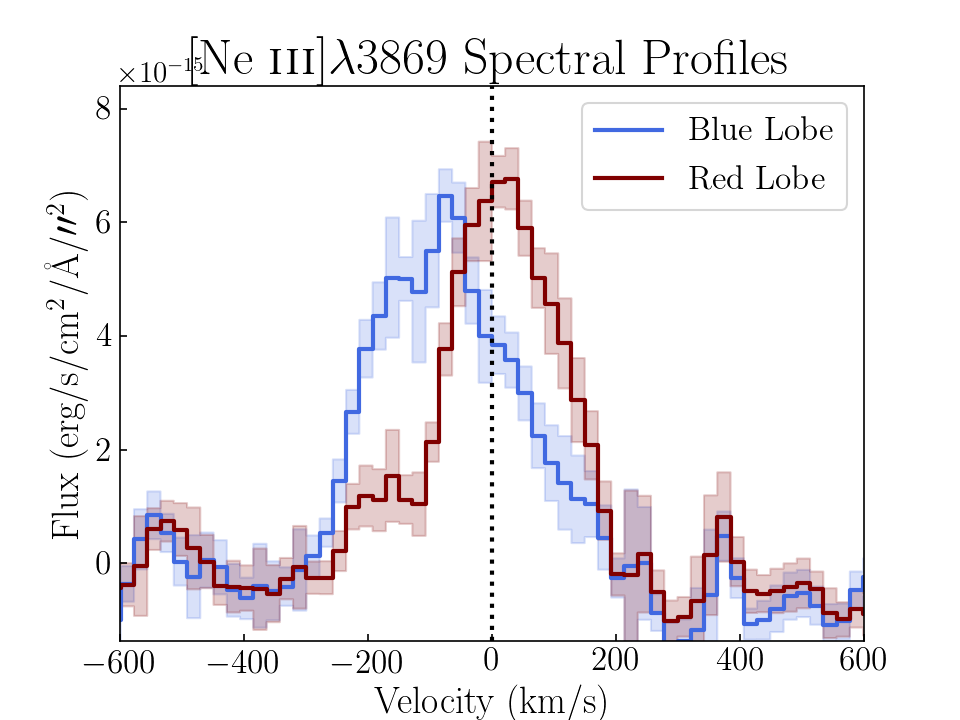} \\
\plotone{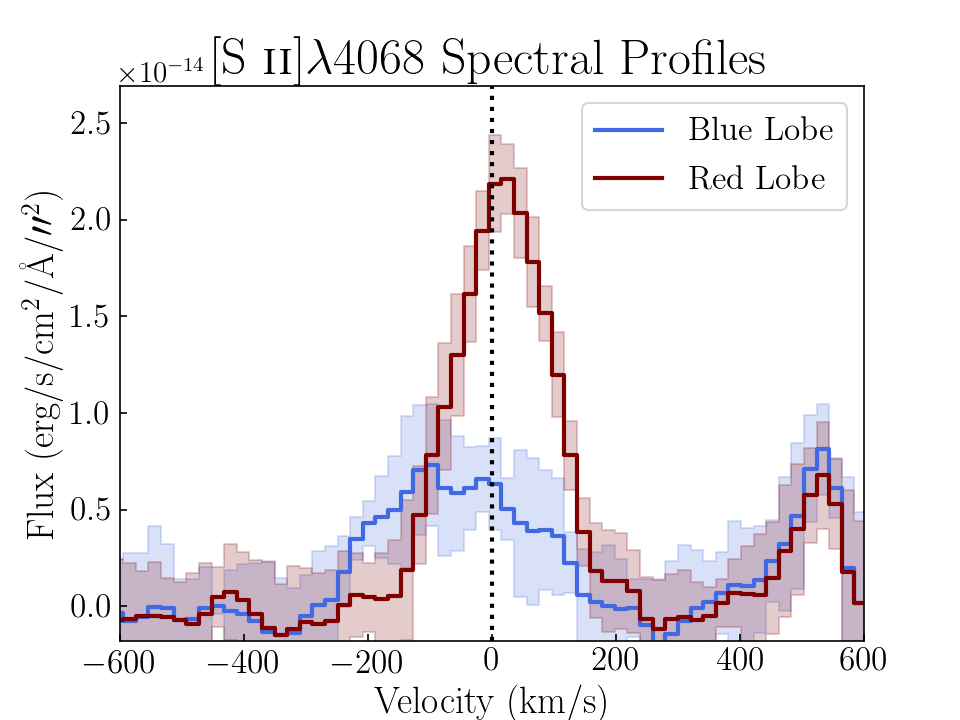}
\caption{One-dimensional spectra of the blueshifted emission (blue) and those of the redshifted emission (red). The profiles were smoothed with a Gaussian of width $\sigma_G=0.8$ cells for better presentations. The peak at $\sim500$ \kms\ in the [\ion{S}{2}] $\lambda4068$ panel is the incomplete emission of [\ion{S}{2}] $\lambda4096$ cut off by the detector edge.} \label{fig:STIS_1d_specprof}
\end{figure}

\begin{figure}
\plotone{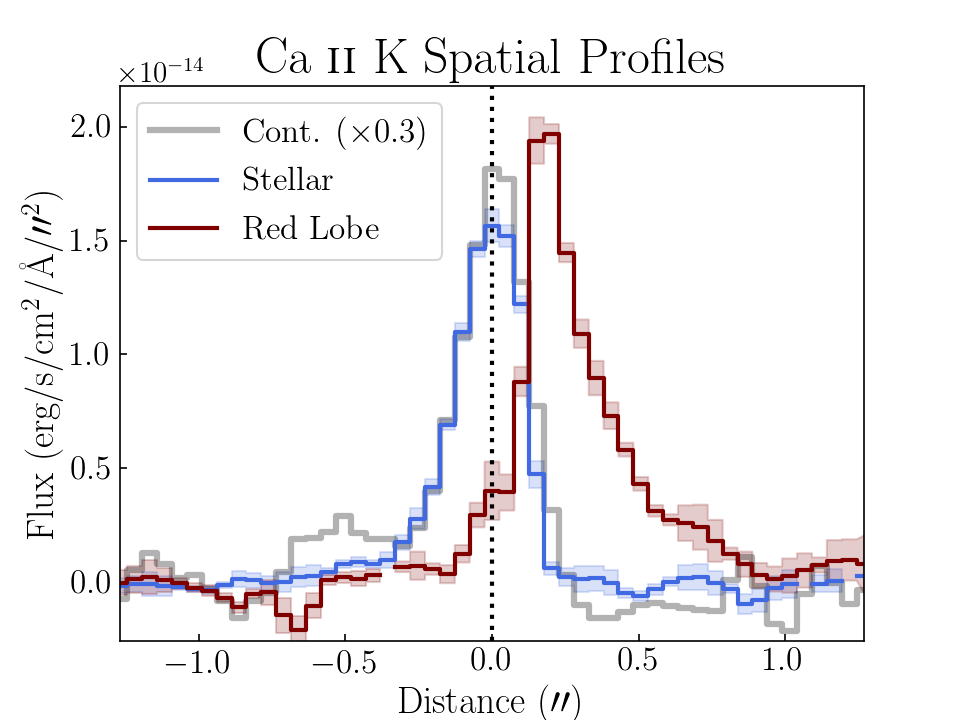} \\
\plotone{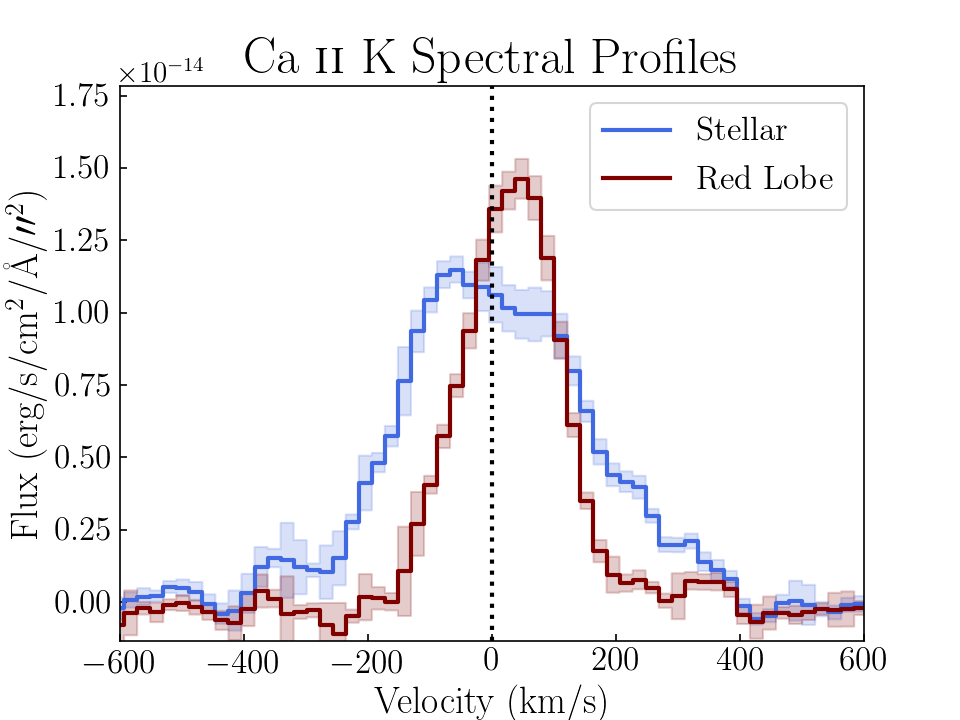}
\caption{One-dimensional spatial (upper) and spectral (lower) profiles of \ion{Ca}{2} K line, decomposed into the stellar component (blue) and the redshifted microjet emission (red). The spatial profile of the continuum, integrated over [$-900,-600$] \kms\ region, is overlaid as a gray line in the upper panel. The profiles were smoothed with a Gaussian of width $\sigma_G=0.8$ cells for better presentations.} \label{fig:STIS_CaIIK_1dprof}
\end{figure}

Figure \ref{fig:STIS_1d_spatprof} shows the spatial line profiles of the blueshifted and redshifted microjets. The intensity peaks of the [\ion{Ne}{3}] and [\ion{S}{2}] lines are within $\sim0\farcs1$ from the nominal stellar center; the [\ion{O}{2}] lines have a peak at $\sim0\farcs4$ in the redshifted microjet and a less evident peak in the blueshifted microjet at $\sim0\farcs1$. The properties of the spatial profiles are consistent with amplitudes fitted from the velocity-decomposed pv diagrams. Figure \ref{fig:STIS_1d_specprof} shows spatially integrated spectral line profiles of each velocity component. Both the high-density lines have a velocity centroid at $\sim -70$ \kms\ in the blueshifted microjet and at $\sim +20$ \kms\ in the redshifted microjet. They also show similar line widths of $\sim 250$ km\,s$^{-1}$ in the blueshifted microjet and $\sim 180$ km\,s$^{-1}$ in the redshifted microjet. The [\ion{O}{2}] lines exhibit similar properties in the redshifted microjet as the high-density tracers, but have larger line width of $\sim -130$ \kms\ than the high-density lines. The permitted \ion{Ca}{2} K line has a velocity centroid of $\sim30$ \kms\ and line width of $\sim180$ \kms, tracing the redshifted microjet as its forbidden-line counterparts. The other component is shown centered at the stellar position and systemic velociy and shows a large line width of $\sim300$ \kms, suggestive of a stellar origin (Figure \ref{fig:STIS_CaIIK_1dprof}). The properties of the integrated line profiles of the forbidden emission are summarized in Table \ref{tab:line_properties}. 

\begin{deluxetable*}{ccccccccccccc}
\tablecaption{Properties of Velocity-Decomposed Emission Lines \label{tab:line_properties}}
\tabletypesize{\footnotesize}
\tablewidth{\textwidth}
\tablehead{ & $d_{\rm cent}$ (\arcsec) & $v_{\rm cent}$ (km\,s$^{-1}$) & $v_{\rm width}$ (km\,s$^{-1}$) & flux (erg\,s$^{-1}$\,cm$^{-2}$)}
\startdata
 & \multicolumn{4}{c}{[\ion{O}{2}] $\lambda3726$} \\
blueshifted & $-0.12\pm0.1$ & $-147\pm19$ & $214\pm31$ & $4.9\pm0.8\times10^{-15}$ \\
redshifted & $0.4\pm0.02$ & $14\pm6$ & $175\pm13$ & $1.1\pm0.1\times10^{-14}$ \\
\hline \hline
 & \multicolumn{4}{c}{[\ion{O}{2}] $\lambda3729$} \\
blueshifted & $-0.16\pm0.01$ & $-123\pm17$ & $276\pm54$ & $3.5\pm0.8\times10^{-15}$ \\
redshifted & $0.45\pm0.05$ & $26\pm10$ & $108\pm22$ & $7.1\pm181\times10^{-15}$ \\
\hline \hline
 & \multicolumn{4}{c}{[\ion{Ne}{3}] $\lambda3869$} \\
blueshifted & $-0.1\pm0.004$ & $-77\pm4$ & $236\pm8$ & $3.5\pm0.2\times10^{-15}$ \\
redshifted & $0.1\pm0.008$ & $20\pm4$ & $194\pm9$ & $3.2\pm0.2\times10^{-15}$ \\
\hline \hline
 & \multicolumn{4}{c}{\ion{Ca}{2} K} \\
stellar & $0.004\pm0.002$ & $2.4\pm2.4$ & $307\pm5$ & $1.1\pm0.03\times10^{-14}$ \\
redshifted & $0.2\pm0.004$ & $30\pm1$ & $174\pm3$ & $9.0\pm0.3\times10^{-15}$ \\
\hline \hline
 & \multicolumn{4}{c}{[\ion{S}{2}] $\lambda4068$} \\
blueshifted & $-0.1\pm0.006$ & $-72\pm15$ & $264\pm40$ & $4.8\pm1.2\times10^{-15}$ \\
redshifted & $0.1\pm0.01$ & $20\pm4$ & $168\pm10$ & $1.4\pm0.1\times10^{-14}$ \\
\enddata
\end{deluxetable*}

\subsection{Physical Conditions of Sz 102 Microjets} \label{subsec:phys_condition}

The [\ion{O}{2}] $\lambda\lambda3729/3726$ ratio is an inverse function of the electron density $n_e$ ranging between 1.5 and 0.35, most sensitive between $10$ to $10^5$ cm$^{-3}$ with only a small dependence on temperature \citep[see, e.g., ][]{Pradhan2006}. From the velocity-decomposed pv diagram [Figure \ref{fig:STIS_2d_spec_vdecomp}(a), (b)], the median value of the [\ion{O}{2}] ratios of the redshifted microjets is $\sim0.44\pm0.23$, corresponding to electron density of $\sim10^4$ cm$^{-3}$. The [\ion{O}{2}] ratio of the blueshifted microjet is poorly determined from the velocity-decomposed pv diagram due to low signal-to-noise. The ratio obtained from the spatially-integrated blueshifted emission (Figure \ref{fig:STIS_1d_spec} and Table \ref{tab:1dspec}) has a low value of $<0.35$, suggesting a higher electron density of $\gtrsim10^5$ cm$^{-3}$.

The [\ion{O}{2}] flux can also be used to check the consistency of the derived electron density by assuming the emitting volume in the jet. The flux can be modeled as coming from an emitting volume $V$ at a distance $d$, $F = \frac{n_u A_{ul} h\nu V}{4 \pi d^2}$. From the observation, the redshifted [\ion{O}{2}] $\lambda3726$ microjet flux is $\sim 10^{-14}$ erg\,s$^{-1}$\,cm$^{-2}$. The emitting volume is assumed to be a cylinder in which its length is approximately the observed extent of the [\ion{O}{2}] microjet of $h\approx0\farcs6$ and its radius is estimated as half the slit width $r\approx0\farcs1$, resulting in the volume of $V\approx7.7\times10^4$ au$^3$ if the distance $d=160$ pc is adopted. Adopting the atomic properties of $A_{ul} = 1.59\times10^{-4}$ and $h\nu = 5.33\times10^{-12}$ erg, the upper population of [\ion{O}{2}] $\lambda3726$ is estimated to be $n_u \approx 0.14$ cm$^{-3}$. We also assume an electron fraction of $x_e \sim 0.1$ and the cosmic abundance of oxygen $x_{\rm O} \sim 4.9\times10^{-4}$ \citep{Asplund2009ARAA}. In order to obtain the optically-thin limit of $n_e \sim10^4$ cm$^{-3}$, it requires either that the emitting volume is transversely unresolved or an ionization fraction of O$^+$ being on the order of $\sim10^{-3}$.

We can obtain an independent constraint on the densities in the redshifted microjet if we assume that the \ion{Ca}{2} emission is cospatial with the forbidden emission lines. The electron number density can be rearranged to be obtained as $n_e = \frac{F}{A_{ul} h\nu}\frac{x_e}{x_{\rm Ca\,II}}\frac{4\pi d^2}{V}$, with the emitting volume $V$ identical to that used in the [\ion{O}{2}] estimation. We adopt $d=160$ pc and the \ion{Ca}{2} K line flux $F\approx10^{-14}$ erg\,s$^{-1}$\,cm$^{-2}$ from the observation, and the atomic properties of \ion{Ca}{2} K $A_{ul} = 1.47\times10^8$ s$^{-1}$ and $h\nu = 5.05\times10^{-12}$ erg. We assume that nearly all the calcium atoms are singly ionized such that the fractional abundance of \ion{Ca}{2} equals to the cosmic abundance of Ca, $x_{\rm Ca\,II} \approx 2.2\times10^{-6}$ \citep{Asplund2009ARAA}, and that the jet is partially ionized, $x_e \approx 0.1$. If the \ion{Ca}{2} K line is optically thin, the required electron density is $n_e \approx 7\times10^{-9}$ cm$^{-3}$. The apparent discrepancy between the inferred electron density using forbidden and permitted emission suggests that the \ion{Ca}{2} emission is very optically thick.

\section{Discussion} \label{sec:discussion}

\subsection{Permitted \ion{Ca}{2} Emission Lines in the Sz 102 Microjets} \label{subsec: permitted_lines}

The \ion{Ca}{2} H and K emission lines are ubiquitous in pre--main-sequence stars as seen from the survey by \citet{Herbig1986}. 
The lines can form in the chromosphere and accretion flows of actively accreting T Tauri Stars \citep{Hartmann1990,KF11}. 
Dense magnetospheric accretion flows may be manifested as broad components and/or redshifted absorption components, while accretion shocks on the star may reveal themselves as narrow emission peaks centered at the stellar velocity \citep[][see also \ion{Ca}{2} near-infrared triplet lines, e.g., \citealt{Muzerolle1998}]{AB00}. 
Spectroscopic studies have also suggested possible association of \ion{Ca}{2} emission with wind and outflow activities. \ion{Ca}{2} H and K, as well as Balmer lines, were detected in bow shocks and knots in HH 1-2 \citep{HR84} and HH 47A \citep{Hartigan1999}. In spectra of some Classical T Tauri Stars, the \ion{Ca}{2} H, K, \ion{Mg}{2} h, k, and \ion{Na}{1} D lines were found to show blueshifted P-Cygni profiles indicative of strong stellar wind activities \citep{Walter1999}. RW Aur is also an example showing clear blueshifted absorption in the \ion{Ca}{2} K line \citep{AB00}.

\begin{figure}
    \centering
    \plotone{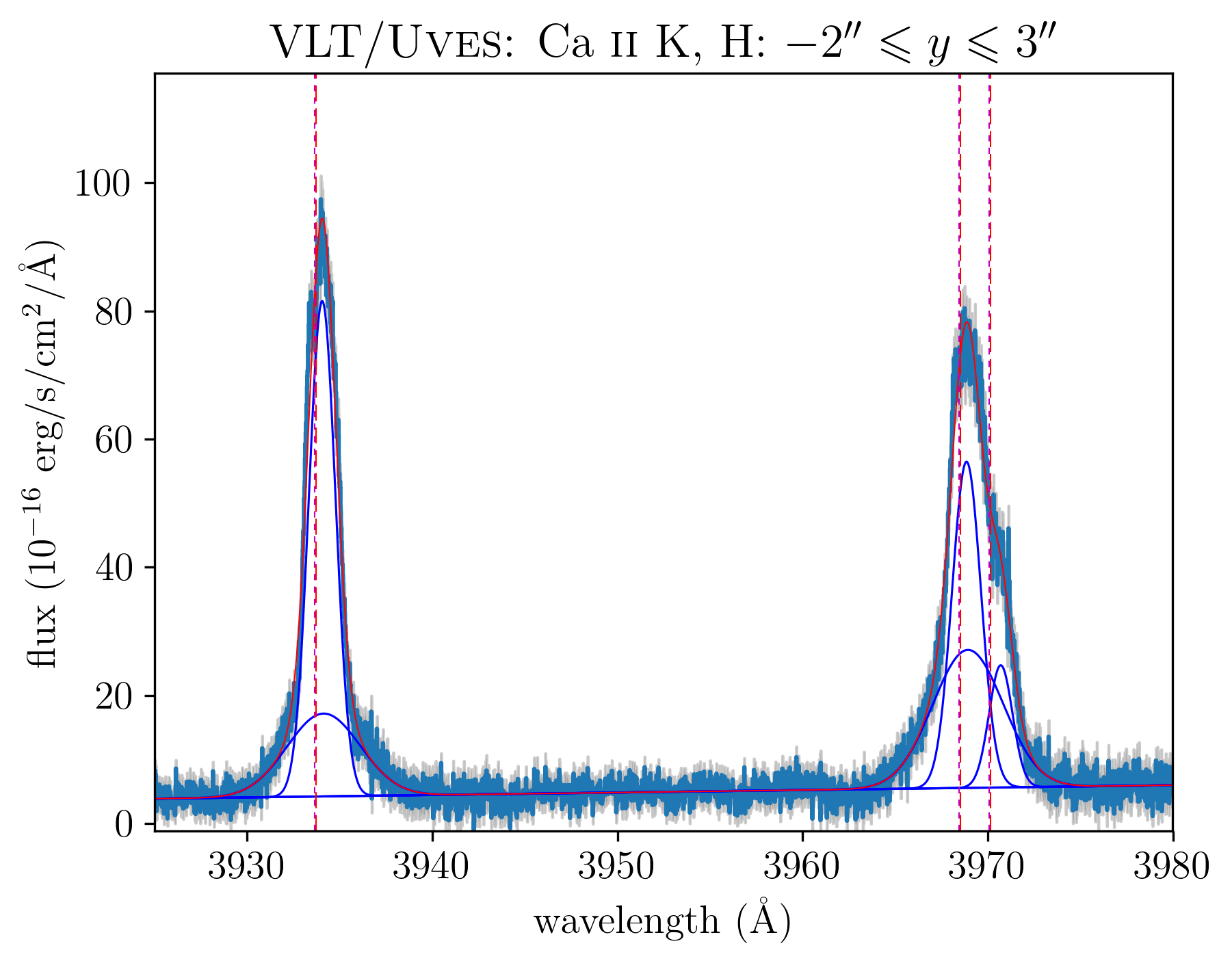}
    \caption{Spatially-integrated (between $-2\arcsec$ and $+3\arcsec$ from the star) spectrum in the \ion{Ca}{2} H and K region obtained from velocity-resolved VLT/{\sc Uves} observations. The vertical dashed lines indicate the systemic line positions of \ion{Ca}{2} K, H, and H$\epsilon$, from left to right. The line profile is fitted by 5 Gaussians representing the narrow and broad components of \ion{Ca}{2} K, the narrow and broad components of \ion{Ca}{2} H, and the blended H$\epsilon$ emission.}
    \label{fig:CaII_UVES_1DSpec}
\end{figure}

One of the novel results from our {\it HST}/STIS spectra is the confirmation of the redshifted \ion{Ca}{2} microjet of Sz 102. Comparisons with data obtained from an earlier epoch suggest the redshifted emission arises from a long-lived microjet. The \ion{Ca}{2} K line profile obtained with VLT/{\sc Uves} was indicative of a composite contribution from the stellar chromosphere and the redshifted microjet \citep{CF10}. The velocity-resolved yet spatially-unresolved VLT/{\sc Uves} line profiles of \ion{Ca}{2} lines consist of a fainter wide component with a line width of $\sim350$ \kms\ and a brighter narrow component with a line width of $\sim130$ \kms\ (Figure \ref{fig:CaII_UVES_1DSpec}). The narrow component has a velocity centroid and line width comparable to the collective properties of the redshifted microjet traced by forbidden emission lines extracted from the same dataset. We decomposed our spatially-resolved {\it HST}/STIS spectrum into the wide and narrow components, with the narrow component spectrally centered at the redshifted jet velocity and spatially tracing the redshifted microjet.

We suggest that the redshifted \ion{Ca}{2} H and K lines trace the high-density regions closest to the jet axis in the cylindrically stratified jets and winds of Sz 102. 
After deconvolving the blended \ion{Ca}{2}~H and H$\epsilon$ lines, the K/H line ratio for the redshifted microjet is $\gtrsim1.4$ (Table \ref{tab:1dspec}), marginally approaching to the optically thick limit of $\sim2$. 
The analysis provided in Sec.\ \ref{subsec:phys_condition} is also consistent with the assumption that \ion{Ca}{2} lines are optically thick in the jet. The \ion{Ca}{2} lines may form in a thin layer surrounding the high-density regions close in to the axis of the microjet, which is transversely unresolved by the slit observation.

\begin{figure}
    \centering
    \plotone{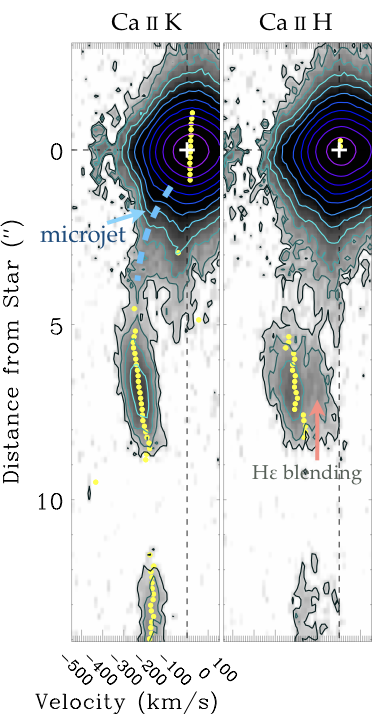}
    \caption{Spatially-resolved pv diagrams of \ion{Ca}{2} K and H emission lines from the DG Tau microjet for the inner $14\arcsec$ obtained by VLT/X-Shooter. Velocity centroids extracted from line profiles at each of the spatial positions are shown in yellow filled circles. The emission close to the star is dominated by stellar chromospheric emission and the region between the star and the knot at $\sim7\arcsec$ shows that the spatially extended microjet traced by forbidden emission is also traced by \ion{Ca}{2} K emission. The velocity and distance along the jet are relative to the systemic velocity and continuum peak position of DG Tau, shown as a white cross.}
    \label{fig:CaII_XSh_DGTau}
\end{figure}

The occurrence of \ion{Ca}{2} emission in the jets of low-mass YSOs may be more prevalent than previously understood.
Figure \ref{fig:CaII_XSh_DGTau} shows the identification of \ion{Ca}{2} K and H emission from the $\sim5\arcsec$ blueshifted microjet of the T-Tauri Star DG Tau, obtained from the VLT/X-Shooter spectra reported in \citet{Liu16}. The bulk of the emission is dominated by the stellar contribution, as expected from the spectrally resolved, but spatially unresolved, spectra \citep[e.g.,][]{AB00}. On the other hand, along the jet axis, \ion{Ca}{2} K emission is visible up to $\sim5\arcsec$ at a velocity of $\sim-150$ \kms, linking the emission from the star and the knot A/B at $\sim7\arcsec$. The H/K line ratio is also between 1 and 2, approaching the optically thick limit. 
Studies of the prevalence of \ion{Ca}{2} jets would require thorough investigations of spatially resolved spectra in order to decompose the jet contributions from the overwhelming stellar contributions.

\subsection{Asymmetries in the Sz 102 Microjets} \label{subsec: asymmetry}

The {\it HST}/STIS spectra, at a spatial resolution of $\sim0\farcs1$, show asymmetries between the blueshifted and redshifted sides of the Sz~102 microjets that can be traced down to the inner tens of au from the outer knots.
Asymmetries in the kinematics are evident in the centroids and widths of the velocity-decomposed pv diagrams and line profiles. The redshifted microjet exhibits an average velocity centroid of $\sim 20$--$30$ \kms, 2 to 3 times slower than the blueshifted microjet; and an average width of $\sim 160$--$190$ \kms, $\sim 1.5$ times smaller than the blueshifted microjet. At larger scales, using long-slit spectroscopy, a blueshifted outer knot at $\sim35\arcsec$ (HH E$_1$) has been reported to be $\sim2$ times faster than its redshifted counterpart at a similar distance (HH W) \citep{GH88}. A similar blue-to-red speed ratio, can be found on an intermediate scale of $\sim200$ au at a $\sim1\arcsec$ resolution in the VLT spectra of near-infrared [\ion{Fe}{2}] line \citep{Coffey2010}, and optical forbidden lines including [\ion{Ne}{3}] \citep{Liu14}, in which a velocity centroid of $\sim+20$ \kms\ and a velocity width of $\sim140$ \kms\ are found for the redshifted jet and $\sim-90$ \kms\ and $\sim190$ \kms\ for the centroid and width of the blueshifted jet. 

The intensity of the blueshifted and redshifted jets also shows asymmetries at various spatial scales. The redshifted jet is brighter and more extended than the blueshifted jet on a scale of $<10\arcsec$ \citep{GH88,WH09,Coffey2010}. This is also evident on the scale of $1\arcsec$ microjets in the $\sim1\arcsec$-resolution VLT optical spectra \citep{Liu14} and in the $\sim0\farcs1$-resolution {\it HST} spectra in this work. This apparent brightness difference and the underlying difference in various line ratios were used to infer the asymmetric physical conditions of the jets launched from two sides of the circumstellar disk. Compared to the redshifted microjet, the blueshifted microjet has a temperature that is $\sim13\%$ higher, the electron density that is $\sim46\%$ higher, and the electron fraction that is a factor of $\sim2$ higher. 

Although the spectral coverage and spectral resolution provided in the current dataset does not allow for full analyses of physical conditions within $1\arcsec$ using various line ratios, spectral properties of the available emission lines are in good agreement with those obtained from the spatially unresolved ground-based spectra. The good agreement of spectral properties between the spatially unresolved and resolved spectra enables us to infer that the asymmetric properties persist up to close to the base of the jet. As demonstrated in the case of RW Aur A, the blueshifted microjet is launched with a higher wind velocity and lower density than the redshifted counterpart, as a result of different magnetocentrifugal mass loading at each side of the disk while maintaining the fundamental conservation laws of momentum flux and magnetic flux \citep{LS12}. Sz~102, with the same patterns of asymmetry, may fit in the same launching scenario.

\subsection{Possible Origins of Ionization in the Sz 102 Microjets} \label{subsec:ionization_origin}

The main ionization sources of the jet may be irradiation from the vicinity of launching region (including soft coronal X-rays, hard X-rays produced by flares, and extreme ultraviolet [EUV] from accretion shocks) and/or shock ionization in the jet. Spatially resolved [\ion{Ne}{3}] emission line observations provide a straightforward test to distinguish the contributions from these potential ionization sources. Irradiation close to the launching region of the jet provides ionization and excitation at the base of the jet and thus gives rise to emission line peak close to the star. On the other hand, shock ionization can only occur at a fast shock exceeding $\sim 100$ \kms, leaving clear kinematic signatures in the emission lines and showing clear offset from the vicinity of the star.

Investigation of possible features of shock ionization in the properties of the Sz 102 microjets does not lead to positive detections of strong shocks on the order of $\sim100$ \kms.
The spatial distributions of the [\ion{Ne}{3}] $\lambda3869$ line, the high-density tracer [\ion{S}{2}] $\lambda4068$, and the permitted \ion{Ca}{2} lines, exhibit a centrally peaked and spatially extended emission pattern. A very bright optical knot along the propagation path of the jet, as required by the shock ionization and excitation, was not detected from the position--velocity profiles of these high-excitation lines. Compared to the VLT/{\sc Uves} spectra obtained $\sim13$ yr ago, if the emission was ionized and excited by a fast shock, the knot would have been able to travel based on the proper motion \citep[$\sim0\farcs37$ yr$^{-1}$, ][]{WH09} down to $\sim 4\arcsec$ from the star, which is not detected within the 52\arcsec-long slit. Spectrally, the velocity structure associated with the strong emission region maintains a relatively constant velocity pattern, in contrast to an abrupt change in velocity centroids and widths expected for a very strong shock of velocity variation, such as the knot B0/B1 structure resolved by {\it HST} in \citet{Maurri2014}. Therefore, it is difficult to apply the shock scenario for the case of Sz 102 for the line emission at the vicinity of the star. 

The spatial distribution of gradually decreasing line intensities and relatively constant line centroids and widths along the flow is consistent with the scenario wherein the jet is ionized at the launching region and that the intensity fades away along the flow due to ion destruction by electron recombination and charge exchange with H atoms. For \ion{Ne}{3}, the dominant destruction channel is electron recombination. 
Assuming physical conditions $T\approx2\times10^4$ K and $n_e\approx6\times10^4$ cm$^{-3}$ appropriate for the redshifted microjet, the recombination timescale of \ion{Ne}{3} is $\sim0.6$ yr
\citep{Liu14}. This allows the ionized neon to freeze in the flow up to a scale of $\sim40$ au, for a jet speed of $\sim250$--$300$ \kms, as deprojected by inclination of $i\gtrsim85^\circ$ \citep[e.g., ][]{CF10}. Considering a nearly edge-on inclination of Sz 102 at a distance of 160 pc, the frozen-in region of ionized neon in the flow corresponds to an microjet extension of $\sim0\farcs24$, consistent with what is observed in the redshifted microjet.

The EUV photons, with energies up to 0.1 keV, may cumulatively reach the ionization states up to \ion{Ne}{5}, but the subsequent electron recombination and H-atom charge exchange in partially ionized medium would quickly transform back to a lower ionization state of \ion{Ne}{2}, possibly accounting for the prevalence of [\ion{Ne}{2}] emission detected in young stellar objects \citep{GNI07}. Moreover, although highly uncertain due to an edge-on orientation, the accretion rate of Sz 102 may be as high as $\approx 4$--$6\times10^{-8}$ $M_\odot$, as inferred from \ion{Ca}{2} $\lambda8662$ \citep{CF10}. There is a high chance that the majority of the EUV photons are absorbed by the accreting material \citep{HG09}, leaving a very low EUV flux to reach the wind encompassing the accretion funnel \citep{Pascucci2014}.

Ionization through keV X-rays with multiple-electron Auger processes that can ionize neon up to \ion{Ne}{6} and \ion{Ne}{7} \citep[e.g., ][]{Muller2017} would be required to sustain the \ion{Ne}{3} abundance in partially ionized circumstellar environment \citep{GNI07}.
The trace of keV X-rays can be found in the energy distribution of Sz 102, which possesses X-ray sources that can be decomposed into a soft and a hard component \citep{Guedel2009}, similar to those jet-driving sources with Two-Absorber X-ray spectrum \citep[][]{Guedel2007}.
The soft component with a temperature fit of $\sim2$ MK \citep{Liu14}, with a luminosity of $\sim 7\times10^{29}$ at a distance of 160 pc, only partially surpassing the K-shell edge of $\sim0.9$ keV \citep{GNI07,Muller2017} and therefore may not dominate the neon ionization at high ionization states unless the cumulative photons of the keV tail effectively contribute to ionization without loss in attenuation.
The hard component, peaked around 1 keV and extended beyond 5 keV energy bins with a luminosity on the order of $\sim10^{28}$ \ergs, is the more plausible ionization source for ionization states higher than \ion{Ne}{3}.

In order for the highly ionized neon to be launched in the wind without significantly reducing the wind speed through strong shocks with shock speed larger than $\sim 100$ \kms, the ionization source that can provide keV X-rays would have to locate extremely close to the driving region or otherwise the X-ray photons would be lost in flux dilution at larger distances.
Estimated from the large proper motion due to an edge-on viewing geometry of Sz 102, the wind speed is around 250 to 300 \kms\ \citep{CF10}. Estimated using spectral model fitting or inferred from observations of gaseous disk kinematics, the mass of the Sz 102 star may be in the range of 0.6 to 2 $M_\odot$ \citep{CF10,Louvet2016}. The ranges of stellar mass and wind speed combine to constrain an uncertainty of the wind launching region of $\sim0.01$ to $\sim0.03$ au from the star. Any source that can contribute keV X-ray photons would have to occur in the region to provide immediate ionization of the jet.

The keV-scale X-rays would most likely originate from flares generated by magnetic reconnection events arising in a star--disk system such as Sz 102.
These flares can generate hard X-rays up to tens to hundreds of MK with luminosities on the order of $10^{30}$ to $10^{32}$ \ergs\ \citep[see, e.g., ][]{Favata2005,Wolk2005}.
In YSOs with disk accretion, more energetic flares and large flares with harder spectra tend to occur more than from those non-accreting or YSOs without disks \citep{BG10}. In the presence of an accretion disk, the reconnection events may occur near the inner edge \citep{Waterfall2019,Waterfall2020}, at the magnetic Y-points at the midplane inside the disk edge, and the helmet streamers above the disk surface \citep{SSGL}. The helmet streamers are located high above the disk and can favor ionization into the inner part of the jet \citep{SGSL,SGLL}. Active young stars are observed to possess large-scale organized magnetic loops elevated above the disk plane on a scale of 5 to 10 stellar radii \citep{Walter1999}. Such magnetic configurations are prone to reconnection when the loops are dragged and twisted by interaction with the surrounding disks.

The frequency of recurrence of the flares in Sz 102 is yet to be constrained due to sparse hard X-ray photons escaping from the system \citep[observations reported in][]{Gondoin2006,Guedel2009}. Surveys of X-ray monitoring observations show that the flares are characterized by their rapid rise and slower decay within $\sim 100$ ks \citep{Wolk2005,Favata2005}. These events may not have been captured by existing X-ray observations of Sz 102 on a time span of $\sim 100$ ks. Historical flares may nonetheless ionize the wind that leads to high abundance of \ion{Ne}{3}. The characteristics of the Sz 102 flares would remain speculative and await further monitoring observations of the [\ion{Ne}{3}] line and X-rays.

\section{Summary}
\label{sec:summary}

We obtained spatially resolved two-dimensional spectra of the Sz 102 microjets with {\it HST}/STIS at a spatial resolution of $\sim0\farcs1$, covering forbidden emission lines of [\ion{Ne}{3}]$\lambda3869$, [\ion{O}{2}]$\lambda\lambda3726+3729$, and [\ion{S}{2}]$\lambda4068$, and the permitted \ion{Ca}{2} H and K emission lines. All the forbidden emission lines trace the redshifted microjet at velocity centroids of $\sim+20$ \kms\ and the blueshifted microjet at $\sim-70$ \kms. 
Permitted \ion{Ca}{2} H and K lines are found to also trace the redshifted microjet, with K/H ratio suggestive of origining from the optically thick region close to the axis. The widths of the lines tracing various density regions are similar, $\sim250$ \kms\ in the blueshifted microjet and $\sim150$ \kms\ in the redshifted microjet, as expected in a cylindrically stratified wide-angle magnetocentrifugal wind such as an X-wind \citep{XWind_V,SSG,SGSL}.

The spatial distribution of the forbidden emission lines revealed by the 
{\it HST}/STIS spectra provides an opportunity to distinguish between different jet ionization mechanisms. 
The [\ion{Ne}{3}] and [\ion{S}{2}] lines, which trace high density gas, are mostly confined within $\sim0\farcs24$ ($\sim40$ au) and show peaks within $\sim0\farcs1$ of the star.
The spatial distribution of [\ion{Ne}{3}] emission is consistent with a jet ionized and heated close to its base and recombining along the flow.
There is no evidence of intensity enhancement nor strong velocity variations in the middle of the jet propagation resulted from strong shock ionization $>100$ \kms\ required for collisionally ionizing neon. The spatial extension of the [\ion{Ne}{3}] emission is consistent with the recombination timescale of \ion{Ne}{3} in a hot ($T\approx10^4$ K) and dense ($n_e\approx10^4$ cm$^{-3}$) jet. 

In order to ionize the jet without altering the kinematics through strong shocks, photoionization in the vicinity of jet launching region is required for immediate ionization without further distance dilution.
Mass estimates in the literature and deprojected wind velocities using proper motion and derived inclination angle combine to constrain the jet launching radius to be on the order of $\lesssim0.03$ au from the star of Sz 102.
In the vicinity close to the star, flares generating keV X-rays may provide the most effective ionization through Auger processes.  
Young active stars are observed to possess large organized magnetic loops up to a few stellar radii. Reconnection events, leading to large flares, may occur in the helmet streamer associated with the large magnetic loops or when the large magnetic loops are twisted by interactions with the inner accretion disks.

\acknowledgments
The authors are grateful to Alfred Glassgold, who suggested the search of this optical [\ion{Ne}{3}] transition. The authors acknowledge grant support from the Ministry of Science and Technology (MoST) of Taiwan through grant 105-2119-M-001-044-MY3, 108-2112-M-001-009-, and 109-2112-M-001-028-, and from grant HST GO 14177-002A to Stony Brook University. This work utilized the high-performance computing resources for Theory in Academia Sinica Institute of Astronomy and Astrophysics (ASIAA). The {\it  HST}/STIS spectra were obtained from the Mikulski Archive for Space Telescopes (MAST) at the Space Telescope Science Institute (STScI) under General Observing Program 14177. STScI is operated by the Association of Universities for Research in Astronomy, Inc., under NASA contract NAS 5-26555. This research has made use of SAO/NASA Astrophysics Data System.

\facility{{\it HST}(STIS)}


\software{Astropy({\tt CCDProc}) \citep{ccdproc_v1.2.0_2016}, pySpecKit \citep{pyspeckit}} 


\bibliographystyle{aasjournal}
\bibliography{Sz102_STIS}

\end{document}